\documentclass[aps,twocolumn,amsmath,amssymb,a4paper]{revtex4}

\usepackage{graphicx}
\usepackage{dcolumn}
\usepackage{bm}
\usepackage{rotating}

\usepackage[utf8]{inputenc}
\usepackage[english]{babel}
\usepackage{bbm}
\usepackage{verbatim}
\usepackage{afterpage}

\usepackage{color}
\definecolor{lightblue}{rgb}{0.2,0.2,0.7}
\definecolor{darkblue}{rgb}{0,0.25,0.5}
\definecolor{redbrown}{rgb}{0.875,0.25,0.125}
\definecolor{darkgreen}{rgb}{0,0.5,0}

\newcommand{\ket}[1]{\ensuremath{\vert #1  \rangle}}

\renewcommand{\b}[1]{\ensuremath{\mathbf{#1}}}
\renewcommand{\H}{\ensuremath{\text{H}}}
\renewcommand{\l}{\ensuremath{\lambda}}

\newcommand{\HF}{\ensuremath{\text{HF}}}

\begin{document}

\title{Double-hybrid density-functional theory applied to molecular crystals}

\author{Kamal Sharkas$^{1,2}$\footnote{Present address: Department of Chemistry, State University of New York at Buffalo, Buffalo, NY 14260-3000, USA}}\email{kamal.sharkas@etu.upmc.fr}
\author{Julien Toulouse$^{1,2}$}\email{julien.toulouse@upmc.fr}
\author{Lorenzo Maschio$^3$}\email{lorenzo.maschio@unito.it}
\author{Bartolomeo Civalleri$^3$}\email{bartolomeo.civalleri@unito.it}
\affiliation{
$^1$Sorbonne Universit\'es, UPMC Univ Paris 06, UMR 7616, Laboratoire de Chimie Th\'eorique, F-75005 Paris, France\\
$^2$CNRS, UMR 7616, Laboratoire de Chimie Th\'eorique, F-75005 Paris, France\\
$^3$Dipartimento di Chimica and Centre of Excellence NIS (Nanostructured Interfaces and Surfaces), Universit{\`a} di Torino, via Giuria 5, I-10125 Torino, Italy}


\date{July 3, 2014}

\begin{abstract}
We test the performance of a number of two- and one-parameter double-hybrid approximations, combining semilocal exchange-correlation density functionals with periodic local second-order M{\o}ller-Plesset (LMP2) perturbation theory, for calculating lattice energies of a set of molecular crystals: urea, formamide, ammonia, and carbon dioxide. All double-hybrid methods perform better on average than the corresponding Kohn-Sham calculations with the same functionals, but generally not better than standard LMP2. The one-parameter double-hybrid approximations based on the PBEsol density functional gives lattice energies per molecule with an accuracy of about 6 kJ/mol, which is similar to the accuracy of LMP2. This conclusion is further verified on molecular dimers and on the hydrogen cyanide crystal.
\end {abstract}

\maketitle

\section{Introduction}

The reliable computational prediction of the lattice energies of molecular crystals is important in materials science~\cite{AakSed-CSR-93,KarAnaFerCaiVicTocFloPri-JPC-07,GouKenLeu-CGD-07,KwoRuiChoMutSchJazGraGun-CM-06}. It requires an accurate treatment of different types of intermolecular interactions, including electrostatics~\cite{CooPriWilLes-JPC-96}, induction~\cite{FerCsoNarAng-JCC-98,WelKarMisStoPri-JCTC-08}, and dispersion interactions~\cite{StoTsu-JPCB-97}. Kohn-Sham density-functional theory (DFT)~\cite{HohKoh-PR-64,KohSha-PR-65} using standard (semi)local density functionals can account for electrostatic and induction interactions in molecular crystals~\cite{MorSid-CEJ-03,ForBroWooVoc-JCP-03,JuXiaCh-IJQC-05}. However, these usual semilocal functionals fail to adequately model the dispersion interactions~\cite{KriPul-CPL-94,HobSpoRes-JCC-95,HonWatSanIitAsp-JPCL-10,ZhaTru-JCTC-07} which can be an important source of attraction even in hydrogen-bonded crystals~\cite{TsuOriHonMik-JPCB-10}. One possibility for improving the prediction of crystal lattice energies is to use empirical~\cite{NeuPer-JPCB-05,LiFen-PR-06,Civalleri:2008p37490,KarDayWelKenLeuNeuPri-JPC-08,UglZicTosCiv-JMC-09,ZicKirCivRam-PCCP-10,SorRic-JPCC-10,BalByrRic-JPCB-11,RecJanPeiBre-JCC-12,OteJoh-JCP-12a,OteJoh-JCP-12} or non-empirical~\cite{DioRydSchLan-PRL-04,ThoCooLiPuzHylLan-PRB-07,ShiWuNakKalVas-JCP-10} dispersion-corrected density functionals.

Another possibility for theoretical prediction of molecular crystal lattice energies is to use wave-function type correlation methods such as second-order M{\o}ller-Plesset perturbation theory (MP2), coupled-cluster singles doubles with perturbative triples [CCSD(T)], or symmetry-adapted perturbation theory (SAPT) on molecular dimers and possibly trimers~\cite{AlfOjaHer-IJQC-96,RosPauFulSto-PRB-99,IkeNagKit-CPL-03,RinShe-CE-08,PodRicSza-PRL-08,HerSch-PRL-08,WenBer-JCTC-11,BygAllMan-JCP-12,SodKecYagHir-PRB-13}. More satisfying is the use of fully periodic wave-function correlation methods such as MP2~\cite{SunBar-JCP-96,AyaKudScu-JCP-01,MarGruPaiKre-JCP-09,GruMarKre-JCP-10,DelHutVan-JCTC-12,DelHutVan-JCTC-13,PisBusCapCasDovMasZicSch-JCP-05}, the random-phase approximation~\cite{LuLiRocGal-PRL-09,LiLuNguGal-JPCA-10,DelHutVan-JCTC-13}, or quantum Monte Carlo~\cite{HonWatSanIitAsp-JPCL-10}. In particular, the periodic local MP2 (LMP2) method as implemented in the CRYSCOR program~\cite{PisBusCapCasDovMasZicSch-JCP-05,Mas-JCTC-11,PisSchCasUsvMasLorErb-PCCP-12} is well suited for weakly bound systems such as rare-gas solids~\cite{Casassa:2008p54747,Halo:2009p30779,Halo:2009p30774} and molecular crystals~\cite{MasUsvSchCiv-JCP-10,MasUsvCiv-CEC-10,MasCivUglGav-JPC-11}.

In the present work, we investigate the performance of a number of double-hybrid (DH) approximations combining semilocal density functionals with periodic LMP2 for prediction of the lattice energies of a set of molecular crystals. The DH approximations have been introduced a few years ago~\cite{Gri-JCP-06} and have proven capable to reach near chemical accuracy for molecular properties such as atomization energies, reaction barrier heights, ionization potentials, and electron affinities. DH approximations using two empirical parameters constructed with the Becke 88 (B) exchange functional~\cite{Bec-PRA-88} and the Lee-Yang-Parr (LYP) correlation functional~\cite{LeeYanPar-PRB-88} have recently been applied to molecular crystals and essentially no improvement over MP2 was found~\cite{DelHutVan-JCTC-12,SanAraOrtOli-JCP-13}. Here, we test the Perdew-Burke-Ernzerhof (PBE)~\cite{PerBurErn-PRL-96} and the PBEsol~\cite{PerRuzCsoVydScuConZhoBur-PRL-08} exchange-correlation functionals in the one-parameter DH scheme recently proposed by some of us~\cite{ShaTouSav-JCP-11}.

The paper is organized as follows. In Section~\ref{sec:theory}, we review the one-parameter DH approximations and the main equations of the periodic LMP2 method, and we indicate the modifications made for the present implementation. After giving computational details on the calculations in Section~\ref{sec:details}, we present and discuss the results obtained with various DH approximations on four molecular crystals: urea, formamide, ammonia, and carbon dioxide. For the best identified DH approximations constructed with the PBEsol functional, we present and discuss further results on molecular dimers and on the hydrogen cyanide crystal. Section~\ref{sec:conclusion} summarizes our conclusions.


\section{Theory}
\label{sec:theory}

\subsection{One-parameter double-hybrid approximations}

After some related earlier work~\cite{ZhaLynTru-JPCA-04,ZhaLynTru-PCCP-05}, Grimme~\cite{Gri-JCP-06} introduced the procedure which is now commonly used in DH approximations. First, a normal self-consistent DFT calculation is carried out using a fraction $a_x$ of Hartree-Fock (HF) exchange $E_{x}^{\HF}$, a fraction $(1-a_x)$ of a semilocal exchange density functional $E_{x}[n]$, and a fraction $(1-a_c)$ of a semilocal correlation density functional $E_{c}[n]$. Subsequently, the MP2 correlation energy $E_{c}^{\text{MP2}}$ calculated using the previously generated orbitals is added with the fraction $a_c$ (see, however, Ref.~\onlinecite{PevHea-JCP-13} for a DH scheme with orbitals optimized in the presence of the MP2 correlation term). The resulting exchange-correlation energy can thus be written as
\begin{eqnarray}
E_{xc}^{\text{DH}} &=& a_x E_{x}^{\HF} + (1-a_x) E_{x}[n] + (1-a_c) E_{c}[n]
\nonumber\\
&& + a_c E_{c}^{\text{MP2}}.
\label{DH}
\end{eqnarray}
Note that in Eq.~(\ref{DH}) we use the notations $E_{x}^{\HF}$ and $E_{c}^{\text{MP2}}$ to designate exchange and correlation energies calculated with the HF and MP2 exchange and correlation energy expressions but the orbitals used in these expressions are not the standard HF orbitals. Many DH approximations relying on this procedure have been developed~\cite{Gri-JCP-06,SchGri-PCCP-06,TarKarSerVuzMar-JPCA-08,KarTarLamSchMar-JPCA-08,SanPer-JCP-09}. For example, the B2-PLYP approximation~\cite{Gri-JCP-06} is obtained by choosing the B exchange functional for $E_{x}[n]$ and the LYP correlation functional for $E_{c}[n]$, and the two empirical parameters $a_x=0.53$ and $a_c=0.27$ as optimized for the G2/97~\cite{CurRagRedPop-JCP-97} subset of heats of formation.

Recently, a rigorous reformulation of the DH approximations was provided~\cite{ShaTouSav-JCP-11} by applying the multideterminant extension of the Kohn-Sham scheme~\cite{TouColSav-PRA-04,AngGerSavTou-PRA-05} to the adiabatic-connection Hamiltonian
\begin{eqnarray}
\hat{H}^\l = \hat{T}+\hat{V}_{\text{ext}}+\l\hat{W}_{ee} + \hat{V}_{\H xc}^{\l}[n],
\label{Hl}
\end{eqnarray}
which links the non-interacting Kohn-Sham Hamiltonian ($\l=0$) to the exact Hamiltonian ($\l=1$). In this expression, $\hat{T}$ is the kinetic energy operator, $\hat{V}_{\text{ext}}$ is a scalar external potential operator (e.g., nuclei-electron), $\hat{W}_{ee}$ is the electron-electron interaction operator, and $\hat{V}_{\H xc}^{\l}[n]$ is the Hartree-exchange-correlation potential operator keeping the one-electron density $n$ constant for all values of the coupling constant $\l$. A nonlinear M{\o}ller-Plesset-like perturbation theory~\cite{AngGerSavTou-PRA-05,FroJen-PRA-08,Ang-PRA-08,ShaTouSav-JCP-11} can then be defined, which when truncated at second order gives the density-scaled one-parameter double-hybrid (DS1DH) approximation 
\begin{eqnarray}
E^{\text{DS1DH},\l}_{xc} &=& \l E_x^{\HF} + (1-\l) E_x[n] 
\nonumber\\
&&+ E_c[n] -\l^2 E_c[n_{1/\l}] + \l^2 E_c^{\text{MP2}},
\label{ExcDS1DH}
\end{eqnarray}
where $E_c[n_{1/\l}]$ is the correlation energy functional evaluated at the scaled density $n_{1/\l}(\b{r})=(1/\l)^3 n(\b{r}/\l)$. Neglecting the density scaling in the correlation functional, $E_c[n_{1/\l}]\approx E_c[n]$, leads to the one-parameter double-hybrid (1DH) approximation
\begin{eqnarray}
E^{\text{1DH},\l}_{xc} &=& \l E_x^{\HF} + (1-\l) E_x[n] + (1-\l^2) E_c[n]
\nonumber\\
&&+ \l^2 E_c^{\text{MP2}}.
\label{Exc1DH}
\end{eqnarray}
Just as for the original DH approximations, in Eqs.~(\ref{ExcDS1DH}) and~(\ref{Exc1DH}), only the first three terms are included in a self-consistent hybrid Kohn-Sham calculation, and the last MP2 term is evaluated with the previously obtained orbitals and added \textit{a posteriori}. In comparison to the original DH approximations, only one empirical parameter needs to be determined. DS1DH and 1DH approximations using the BLYP and PBE exchange-correlation functionals have been constructed and the optimal parameter was found to be about $\l\approx 0.65$ or $0.70$ for atomization energies and reaction barrier heights of molecular systems~\cite{ShaTouSav-JCP-11}.

\subsection{Periodic local MP2}

We now briefly review the main equations of the periodic local MP2 method that we use and indicate the required modifications for implementing the one-parameter DH approximations. The implementation of two-parameter DH approximations requires obvious similar modifications.

The first-order perturbative correction to the HF wave function is written as~\cite{PisMasCasHalSchUsv-JCC-08}
\begin{eqnarray}
\ket{\Psi^{(1)}} &=& \frac {1}{2} \sum_{(\b i \b j) \in P} \sum_{(\b a \b b) \in [\b i \b j]} T^{\b i \b j}_{\b a \b b} \ket{\Phi^{\b a \b b}_{\b i \b j}},\label{PLMP2WF}
\end{eqnarray}
where $\Phi^{\b a \b b}_{\b i \b j}$ are doubly excited spin-adapted configurations and $T^{\b i \b j}_{\b a \b b}$ are the corresponding amplitudes. In this expression, the labels ($\b i$,$\b j$) refer to pairs of occupied Wannier functions (WFs) taken from a truncated list $P$, in which the first WF $\b i$ is located in the reference unit cell and the second WF $\b j$ is restricted within a given distance to the first WF $\b i$. Note that bold indices here combine the index within the unit cell and the translation (lattice) vector: $\b i$=$i{\mathcal I}$, in the notation of Ref.~\onlinecite{PisMasCasHalSchUsv-JCC-08}. The labels $(\b a$, $\b b)$ refer to pairs of mutually non-orthogonal virtual projected atomic orbitals (PAOs) and the sum is restricted to the pair-domain $[\b i \b j]$ of PAOs which are spatially close to at least one of the WF $\b i$ or $\b j$. This truncation of the virtual space makes the computational cost of the LMP2 method scale linearly with the supercell size.

The double excitation amplitudes $T^{\b i \b j}_{\b a \b b}$ are obtained by solving the following system of linear equations~\cite{PulSae-TCA-86,PisBusCapCasDovMasZicSch-JCP-05,PisMasCasHalSchUsv-JCC-08}
\begin{align} 
K^{\b i \b j}_{\b a \b b} &+ \sum_{(\b c \b d) \in [\b i \b j]} \Bigg \{ F_{\b a \b c} T^{\b i \b j}_{\b c \b d} S_{\b d \b b} + S_{\b a \b c} T^{\b i \b j}_{\b c \b d} F_{\b d \b b} \Bigg \}
\nonumber\\ 
&- \sum_{(\b c \b d) \in u[\b j]} S_{\b a \b c}  \sum_{\b k \,{\it near}\, \b j} F_{\b i \b k} T^{\b k \b j}_{\b c \b d} S_{\b d \b b} 
\nonumber\\ 
&- \sum_{(\b c \b d) \in u[\b i]} S_{\b a \b c} \sum_{\b k \,{\it near}\, \b i}  T^{\b i \b k}_{\b c \b d} F_{\b k \b j} S_{\b d \b b}  =0,
\label{Amp}
\end{align}
where $u[\b i]$ ($u[\b j]$) stands for the union of all $[\b i \b k]$ ($[\b j \b k]$) domains in the $\b k$ summations, which in turn are limited to $\b k$ elements spatially close to $\b i$ (or $\b j$). In Eq. (\ref{Amp}), $K^{\b i \b j}_{\b a \b b}=(\b i \b a|\b j \b b)$ are the two-electron exchange integrals, $S_{\b d \b b}$ is the overlap between PAOs, and $F_{\b i \b k}$ and $F_{\b d \b b}$ are elements of the Fock matrix in WF and PAO basis, which is obtained by transformation of the Fock matrix in the atomic orbital (AO) basis 
\begin{eqnarray}
F_{\mu \nu} &=& h_{\mu \nu} + J_{\mu \nu}+ K_{\mu \nu}, 
\label{FockMat}
\end{eqnarray}
where $h_{\mu \nu}$, $J_{\mu \nu}$, and $K_{\mu \nu}$ are the one-electron Hamiltonian, Coulomb, and exchange matrices, respectively. For closed-shell systems, $J_{\mu \nu} = \sum_{\rho \sigma} P_{\rho \sigma} (\mu \nu| \sigma \rho)$ and $K_{\mu \nu} = (-1/2) \sum_{\rho \sigma} P_{\rho \sigma} (\mu \rho| \sigma \nu)$ where $P_{\rho \sigma}$ is the density matrix and $(\mu \nu| \sigma \rho)$ are the two-electron integrals. In the local orbital basis, the occupied-occupied and virtual-virtual blocks of the Fock matrix are not diagonal, which means that Eq.~(\ref{Amp}) has to be solved iteratively for the amplitudes $T^{\b i \b j}_{\b a \b b}$. When the convergence is reached, the LMP2 correlation energy per unit cell is given as
\begin{eqnarray}
E_c^{\mathrm{LMP2}} = \sum_{\b i \b j \in P} \sum_{(\b a \b b) \in [\b i \b j]} K^{\b i \b j}_{\b a \b b}(2T^{\b i \b j}_{\b a \b b}-T^{\b i \b j}_{\b b \b a}).
\label{EcMP2PerCell}
\end{eqnarray} 

For the one-parameter DH approximations, the Fock matrix of Eq.~(\ref{FockMat}) is replaced by 
\begin{eqnarray}
F^{\mathrm{hybrid}}_{\mu \nu} &=& h_{\mu \nu} + J_{\mu \nu}+ \l K_{\mu \nu} + V^{\l}_{xc,\mu \nu},
\label{KSMat}
\end{eqnarray}
with the scaled exchange matrix $\l K_{\mu \nu}$ and the exchange-correlation potential matrix $V^{\l}_{xc,\mu \nu}$ corresponding to the density functional used. Note that, because of self consistency, the density matrix $P_{\rho \sigma}$ in $J_{\mu \nu}$ and $K_{\mu \nu}$ also implicitly depends on $\l$. Eq.~(\ref{KSMat}) is the essential modification to be made in the program for the LMP2 calculation. Indeed, one can see that using the scaled interaction $\l \hat{W}_{ee}$ of Eq.~(\ref{Hl}) corresponds to scaling the two-electron exchange integrals, $K^{\b i \b j}_{\b a \b b} \to \l K^{\b i \b j}_{\b a \b b}$, in Eq.~(\ref{Amp}), implying the scaling of the amplitudes $T^{\b i \b j}_{\b a \b b} \to \l T^{\b i \b j}_{\b a \b b}$. Using Eq.~(\ref{EcMP2PerCell}), it means that we just need to scale the LMP2 correlation energy by $\l^2$
\begin{eqnarray}
E_c^{\mathrm{LMP2}} \to \l^2 E_c^{\mathrm{LMP2}},
\end{eqnarray}
as it was already indicated in Eqs.~(\ref{ExcDS1DH}) and~(\ref{Exc1DH}).

There is an additional point to consider when using the dual-basis set scheme~\cite{SteDiSShaKonHea-JCP-06,DiSSteHea-MP-07}. The reliable description of the correlated wave functions needs the use of rather large basis sets and especially with diffuse functions when treating weakly bound systems.  However, such basis sets frequently lead to linear-dependency problems in the periodic self-consistent-field calculation (SCF) that precedes the LMP2 step. The dual-basis set scheme  helps overcome these difficulties by using a smaller basis for the SCF calculation and additional basis functions for the LMP2 calculation. In this scheme, Brillouin's theorem does not apply in the LMP2 calculation and hence there is a first-order energy contribution due to single excitations~\cite{PisMasCasHalSchUsv-JCC-08,UsvMasPisSch-ZPC-10} that is conveniently evaluated in reciprocal space in CRYSCOR.
For the case of the one-parameter DH approximations, since the two-electron integrals do not explicit appear in the 
singles equations,
the evaluation of the single excitation contribution do not require other modifications than simply using the dual-basis version of the Fock matrix in Eq.~(\ref{KSMat}) without any additional scaling.
We note, however, that because the one-parameter DH approximations are based on a nonlinear Rayleigh-Schr\"odinger perturbation theory~\cite{AngGerSavTou-PRA-05,Ang-PRA-08,FroJen-PRA-08}, when Brillouin's theorem does not apply, there is in principle an additional single-excitation contribution to the second-order correlation energy coming from the second-order functional derivative of the Hartree-exchange-correlation energy, but we neglect this additional contribution in this work.

\section{Computational details}
\label{sec:details}

The one-parameter DS1DH and 1DH approximations using the BLYP, PBE, and PBEsol functionals, as well as the two-parameter DH approximations, B2-PLYP ($a_x=0.53$, $a_c=0.27$)~\cite{Gri-JCP-06}, B2GP-PLYP ($a_x=0.65$, $a_c=0.36$)~\cite{KarTarLamSchMar-JPCA-08}, and mPW2-PLYP ($a_x=0.55$, $a_c=0.25$)~\cite{SchGri-PCCP-06}, have been implemented in a development version of the CRYSTAL09~\cite{Cry-PROG-09} and CRYSCOR09~\cite{Mas-JCTC-11,PisSchCasUsvMasLorErb-PCCP-12} suite of programs. For the DS1DH approximations, the expressions of the density-scaled correlation energy $E_c[n_{1/\l}]$ and the corresponding potential are given in the appendix of Ref.~\onlinecite{ShaTouSav-JCP-11}. The computational cost of the DH approximations is essentially the same as the one of LMP2.

As our main benchmark set, we take four molecular crystals: urea CO(NH$_2$)$_2$, formamide HCONH$_2$, ammonia NH$_{3}$ and carbon dioxide CO$_{2}$. Although this set is statistically small, it includes systems ranging from a dispersion-dominated crystal (carbon dioxide) to structures with hydrogen bonds of varying strengths (urea, formamide, ammonia). Lattice energies span a range between 28 kJ/mol and 103 kJ/mol. Interestingly, urea and formamide show a 3D and 2D network of hydrogen bonds, respectively, and then sheets in formamide interact through weaker CH...O intermolecular contacts. The experimental crystal structures were used~\cite{All-AC-02,SimPet-AC-80,KiePenBreClo-JCP-87,VoiTolMan-JLTP-71} and the C$-$H, N$-$H, and O$-$H  bond distances were rescaled to 1.08, 1.00, and 1.00 {\AA} respectively, following geometrical procedures proposed in literature~\cite{WilPriLesCat-JCC-95,Gav-MP-08}. The truncation tolerances of lattice sums for one- and two-electron integrals were set to 7 7 7 12 40 (TOLINTEG parameters~\cite{Cry-PROG-09}). A grid consisting of 75 radial points and up to 974 angular points was used in evaluating the exchange-correlation functional. The shrinking factors were set to 4 for the $\b k$ points grid to sample the irreducible Brillouin zone. Each calculation is done in two steps: a periodic SCF hybrid calculation is first performed using CRYSTAL09, and then the periodic LMP2 correlation energy is calculated using CRYSCOR09 and added to the SCF energy after multiplication by the proper scaling factor. 

For urea and formamide, we use the polarized split-valence double-zeta Gaussian basis set 6-31G(d,p)~\cite{HarPop-TCA-73} in the SCF calculations~\cite{PerPanBlaZha-CPL-04}. This basis is then augmented by polarization functions with small exponents ($d$ functions for C, N and O atoms, $p$ functions for H atoms, taken from the aug-cc-pVDZ basis set~\cite{Dun-JCP-89}), and the resulting basis set is denoted by p-aug-6-31G(d,p)~\cite{MasCivUglGav-JPC-11}. The virtual space of the extended basis set is employed in the subsequent LMP2 calculation. When we use this dual-basis set technique, the contribution of single excitations to the LMP2 correlation energy is evaluated and added. In practice, the dual-basis matrices are obtained through a non-self-consistent SCF (GUESDUAL~\cite{Cry-PROG-09}) which uses the density matrix from a previous SCF run to allocate the additional basis functions. For ammonia and carbon dioxide, we use the p-aug-6-31G(d,p) basis set for both the SCF and LMP2 calculations to avoid the significant contribution of single excitations for these systems ($\sim$ 20$\%$ of the calculated lattice energy). For urea and formamide where we use the dual-basis set technique, the contribution of single excitations does not exceed 3$\%$ of the calculated lattice energy. 

Core electrons are kept frozen in all LMP2 calculations. The occupied valence orbital space is spanned by the localized~\cite{ZicDovSau-JCP-05} symmetry-adapted~\cite{CasZicPia-TCA-06} mutually orthogonal Wannier functions (WFs) supplied by the PROPERTIES module of CRYSTAL. The virtual orbital space is spanned by mutually nonorthogonal PAOs, which are constructed by projecting the individual AO basis functions on the virtual space~\cite{PulSae-TCA-86}. The explicit computations cover WF pairs up to distance $d_{\b i \b j}=12$ {\AA}, where the two-electron repulsion integrals were evaluated via the density fitting periodic (DFP) scheme~\cite{MasUsv-PRB-08} for $d_{\b i \b j} \le 8$ {\AA} and via the multipolar approximation for $8 \le d_{\b i \b j} \le 12$ {\AA}. The contribution of the WF pairs with $d_{\b i \b j} \ge 12 $ {\AA} to the correlation energy was estimated through the Lennard-Jones extrapolation technique. Excitation PAO domains have been restricted to the molecular units.

Single-point, static (i.e. without thermal and vibrational zero-point effects) energies are computed to evaluate the counterpoise-corrected lattice energy $E^{\mathrm{CP}}_{\mathrm{LE}}(V)$ per molecule at a given volume $V$ of the unit cell
\begin{eqnarray}
E^{\mathrm{CP}}_{\mathrm{LE}} (V) &=& E_{\mathrm{bulk}}(V)/Z - E^{\mathrm{[gas]}}_{\mathrm{mol}} - E^{\mathrm{[bulk]}}_{\mathrm{mol+ghosts}} (V)
\nonumber\\
&& + E^{\mathrm{[bulk]}}_{\mathrm{mol}}(V),
\label{CPLE_1}
\end{eqnarray}
where $Z$ is the number of the molecular units in the unit cell, $E_{\mathrm{bulk}}(V)$ the total energy of the bulk system (per cell), and $E^{\mathrm{[bulk]}}_{\mathrm{mol+ghosts}}(V)$ and $E^{\mathrm{[bulk]}}_{\mathrm{mol}}(V)$ the total energy of the molecule in the crystalline bulk geometry with and without ghost functions, respectively, at a given cell volume. At least 50 ghost atoms surrounding the central molecule were used. The ghost functions, in the standard Boys-Bernardi counterpoise scheme~\cite{BoyBer-MP-70}, are supposed to eliminate the inconsistency between the finite basis sets used in the molecular and bulk calculations to obtain basis set superposition error (BSSE) free lattice energies. The total energy of an isolated molecule in the gas phase, $E^{\mathrm{[gas]}}_{\mathrm{mol}}$, was computed at the experimental geometry~\cite{Nis-BOOK-13}. When the contribution coming from the relaxation of the geometry of the molecule from the bulk to the gas phase, $\Delta E^{\mathrm{relax}}_{\mathrm{mol}} = E^{\mathrm{[bulk]}}_{\mathrm{mol}} - E^{\mathrm{[gas]}}_{\mathrm{mol}}$, is neglected, Eq.~(\ref{CPLE_1}) reduces to $E^{\mathrm{CP}}_{\mathrm{LE}} (V) = E_{\mathrm{bulk}}(V)/Z - E^{\mathrm{[bulk]}}_{\mathrm{mol+ghosts}}(V)$. According to Refs.~\onlinecite{MasUsvCiv-CEC-10} and~\onlinecite{DelHutVan-JCTC-12}, this relaxation contribution is indeed negligible for ammonia and carbon dioxide, but it can be of the order of $-10$ kJ/mol for urea at the MP2 level (and similarly for formamide). So in this work we do not neglect the relaxation contribution and use Eq.~(\ref{CPLE_1}).

The calculated lattice energies of crystalline urea, formamide, ammonia and carbon dioxide are compared to the experimental sublimation enthalpies, $\Delta H_{\mathrm{sub}}(298.15\mathrm{K})$, after removal of thermal and vibrational zero-point effects~\cite{ReiTka-JCP-13}.
The average experimental uncertainty on sublimation enthalpies is $\pm 4.9$ kJ/mol~\cite{Chi-NS-03,OteJoh-JCP-12} which should be taken as an estimation of the precision limit of the reference data. 

To further assess the performance of the best identified DH approximations, DS1DH-PBEsol and 1DH-PBEsol, we also perform calculations on the hydrogen cyanide crystal, and on molecular dimers (with the MOLPRO program~\cite{Molproshort-PROG-12}) using Dunning basis sets~\cite{Dun-JCP-89}. Computational details for these calculations are given in Sections~\ref{sec:dimers} and~\ref{sec:hcn}.

\section{Results and discussion}

\subsection{Estimation of the basis-size error}

\begingroup
\begin{table*}[t] \centering
\caption{Counterpoise-corrected lattice energies per molecule (in kJ/mol) of the urea, formamide, ammonia, and carbon dioxide crystals, calculated with the LMP2 method with three different basis sets.}
\label{tab:LE2}
\begin{tabular}{lcccccccccc}
\hline
\hline
Basis set          &&       && CO(NH$_2$)$_2$        && HCONH$_2$             && NH$_{3}$    && CO$_{2}$     \\
\hline
6-31G(d,p)         &&       && 78.84      && 56.33       && 25.94 && 18.30     \\
p-aug-6-31G(d,p)   &&       && 92.66$^a$  && 69.26$^a$   && 32.08 && 27.80        \\
p-aug-6-311G(d,p) &&        && 96.94$^a$  && 68.39$^a$   && 31.22 && 33.32      \\
\hline
\hline
\multicolumn{11}{l}{$^a$With dual-basis set technique: 6-31G(d,p) basis for SCF. }\\
\end{tabular}
\end{table*}
\endgroup

We start by estimating the error due to the basis set, using LMP2 since it is the most basis-size dependent method. Table~\ref{tab:LE2} shows the LMP2 lattice energies calculated using the 6-31G(d,p), p-aug-6-31G(d,p), and p-aug-6-311G(d,p) basis sets. The latter triple-zeta basis set has been tailored according to the same technique (outlined in the Computational Details section) used for p-aug-6-31G(d,p). The large difference between the values calculated with the 6-31G(d,p) basis set and those calculated with the p-aug-6-31G(d,p) basis set shows that the augmentation of the 6-31G(d,p) basis set with low-exponent polarization functions, which act as diffuse functions, is mandatory for a correct description of the lattice energies. For the formamide and ammonia crystals, the differences in the LMP2 lattice energies calculated with the p-aug-6-311G(d,p) and the p-aug-6-31G(d,p) basis sets are quite small, less than 1.0 kJ/mol. The largest differences in the LMP2 lattice energies between these two basis sets are obtained for the urea and carbon dioxide crystals with 4.28 and 5.52 kJ/mol, respectively. These largest variations of the lattice energies are thus similar to the average experimental uncertainty on the sublimation enthalpies. This justifies the use of the p-aug-6-31G(d,p) basis set in the following. 

The basis-set dependence of the DH approximations increases with the fraction of MP2 correlation $a_c$. In practice, due to the relatively large values of $a_c$ used (usually between about $0.3$ and $0.6$), it turns out that they do not have a substantially smaller basis-size dependence than standard MP2 (see, e.g., Refs.~\onlinecite{SchGri-PCCP-06,ShaTouSav-JCP-11,KarMar-JCP-11,ChuChe-JCC-11}). Therefore, the above estimates of the basis-size error on the lattice energies apply as well to the DH approximations.

\subsection{Comparison of the methods as a function of $\l$}
\label{sec:comparison}

\begin{figure*}
\begin{center}
\includegraphics[scale=0.38,angle=-90]{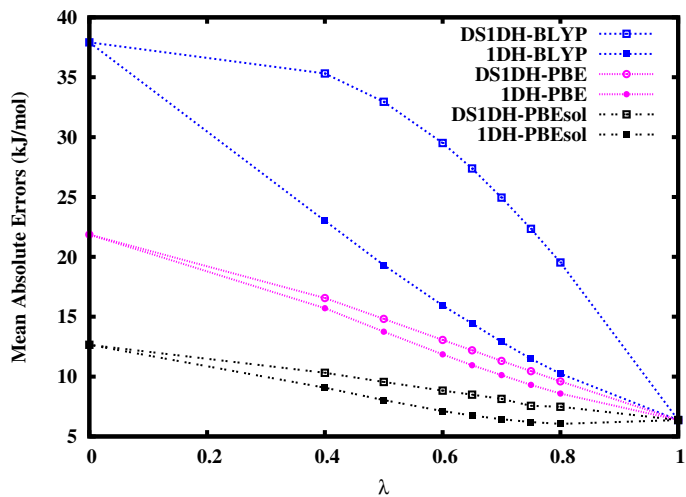}
\end{center}
\caption{MAEs on the counterpoise-corrected lattice energies per molecule (in kJ/mol) of the urea, formamide, ammonia, and carbon dioxide crystals, as functions of the parameter $\l$ for the DS1DH and 1DH approximations with the BLYP, PBE, and PBEsol exchange-correlation density functionals.}
\label{MAELE}
\end{figure*}


To have a first global view of the performance of the different one-parameter DH approximations and the dependence on the parameter $\l$, we show in Fig.~\ref{MAELE} mean absolute errors (MAEs) on the four lattice energies, calculated using Eq.~(\ref{CPLE_1}), as a function of $\l$ for the DS1DH and 1DH methods using the BLYP, PBE, and PBEsol density functionals. For $\l=0$, each method reduces to a standard periodic Kohn-Sham calculation with the corresponding density functional. For $\l=1$, all methods reduce to a standard periodic LMP2 calculation. Kohn-Sham BLYP and PBE calculations at $\l=0$ give much larger MAEs (about 38 kJ/mol and 22 kJ/mol, respectively) than LMP2 (about 6.4 kJ/mol). The DS1DH and 1DH approximations using the BLYP and PBE functionals inherit the bad performance of these functionals and always give larger MAEs than standard LMP2 for all $\l < 1$. In contrast, the PBEsol functional (which is a modified version of the PBE functional which improves the description of solids by restoring the density-gradient expansion of the exchange energy) gives a MAE at $\l=0$ of about 13 kJ/mol. The 1DH-PBEsol approximation gives a slight minimum at around $\l=0.80$. Note that neglecting density scaling in the LYP correlation functional, {\it i.e.} going from DS1DH-BLYP to 1DH-BLYP, significantly reduces the MAEs, which is similar to what was observed for atomization energies of molecular systems~\cite{ShaTouSav-JCP-11}. However, neglecting density scaling in the PBE and PBEsol density functionals only marginally decreases the MAEs on the lattice energies of the molecular crystals considered here, whereas it was found that molecular atomization energies were significantly deteriorated when neglecting density scaling in PBE~\cite{ShaTouSav-JCP-11}. The understanding of the effects of density scaling with different functionals for calculating diverse properties requires further study.

\subsection{Comparison of the methods for each system}

\begingroup
\squeezetable
\begin{table*}[t] \centering
\caption{Counterpoise-corrected lattice energies per molecule (in kJ/mol) of the urea, formamide, ammonia and carbon dioxide crystals, calculated by several methods. For the DS1DH-BLYP, 1DH-BLYP, DS1DH-PBE, and 1DH-PBE double-hybrid approximations, we use the values of $\l$ which were previously optimized in Ref.~\onlinecite{ShaTouSav-JCP-11}. For the DS1DH-PBEsol and 1DH-PBEsol double-hybrid approximations, we use a value of $\l=0.80$ which roughly minimizes the MAE of 1DH-PBEsol for this set of molecular crystals. All calculations were carried out with experimental geometries. For each method, the value with the largest error is indicated in boldface. Mean absolute percentage errors (MA\%Es) are also given.}
\label{tab:LE}
\begin{tabular}{lcccccccccc}
\hline
\hline
Method      & Parameter $\lambda$ && CO(NH$_2$)$_2$$^a$ & HCONH$_2$$^a$ & NH$_{3}$$^b$ &CO$_{2}$$^b$ && MAE  & MA\%E \\
\hline
HF           &                    &&66.02        &\b{37.54}    &6.70    & 1.61     && 33.86  & 66.1 \\
BLYP         &                    &&\b{49.79}    &32.77        &14.60   &-1.49     && 37.91  & 69.0   \\
PBE          &                    &&\b{71.23}    &50.75        &28.57   &9.29      && 21.86  & 39.2  \\
PBEsol       &                    && 86.12       &\b{62.23}    &38.21   &12.15     && 12.66  & 24.3    \\
LMP2         &                    && 92.66       &\b{69.26}    &32.08   & 27.80    && 6.37   &  9.5  \\
\\
Global hybrids \\
B3LYP        &                    &&\b{65.28}    &43.27       &19.00   &3.86      &&28.98    & 54.3\\
PBE0         &                    &&79.56        &\b{54.26}   &27.22   &8.83      && 19.36   & 37.4\\
\\
Empirical dispersion corrected\\
PBE--D3         &                    &&101.72        &76.57   &\b{42.90}   &26.13      &&   2.59 & 4.2\\
B3LYP--D3       &                    && \b{108.70}  &80.54       &39.27   &29.16      &&   2.85 & 4.6\\
\\                                                       
Double hybrids \\
B2-PLYP  &                        &&\b{74.44}   &55.59        &25.30  &15.12       &&19.22   & 34.0\\
B2GP-PLYP   &                     &&84.00      &\b{60.28}    &27.42  &18.49       &&14.28    & 25.8\\
mPW2-PLYP   &                     &&\b{83.56}  &62.49        &29.70  &21.12       &&12.61    & 21.3\\
DS1DH-BLYP  &    $\lambda=0.70$   &&\b{69.36}  &48.95        &19.34  &9.84        &&24.95    & 46.0 \\
1DH-BLYP    &    $\lambda=0.65$   &&\b{82.55}  &60.46        &27.52  &19.04       &&14.43    & 25.5\\
DS1DH-PBE   &    $\lambda=0.65$   &&\b{85.85}  &63.36        &30.20  &19.14       &&12.19    & 21.9  \\
1DH-PBE     &    $\lambda=0.65$   &&\b{87.82}  &64.88        &31.29  &19.49       &&10.95    & 19.9\\
DS1DH-PBEsol &   $\lambda=0.80$   &&\b{92.02}  &\b{68.72}    &32.96  &23.67       &&7.48     & 12.9   \\
1DH-PBEsol  &    $\lambda=0.80$   &&94.21      &\b{70.45}    &34.23  &24.21       &&6.05     & 10.5 \\
\\
Best estimate$^c$ &               &&102.50     &79.20        &37.20  &28.40       &&       \\
\hline
\hline
\multicolumn{11}{l}{$^a$With dual-basis set technique: 6-31G(d,p) basis for SCF and p-aug-6-31G(d,p) basis for LMP2.}\\
\multicolumn{11}{l}{$^b$With p-aug-6-31G(d,p) basis for both SCF and LMP2 calculations.}\\
\multicolumn{11}{l}{$^c$From Ref.~\onlinecite{ReiTka-JCP-13}.}
\end{tabular}
\end{table*}
\endgroup

Table~\ref{tab:LE} reports the lattice energies of the four studied molecular crystals calculated by HF, LMP2, Kohn-Sham with different functionals (including empirical dispersion corrected ones~\cite{GriAntEhrKri-JCP-10}), and various DH methods. For DS1DH-BLYP, 1DH-BLYP, DS1DH-PBE, and 1DH-PBE, we use the values of $\l$ previously optimized on a set of atomization energies and reaction barrier heights of molecular systems~\cite{ShaTouSav-JCP-11} ($\l=0.65$ or $0.70$ depending on the DH method considered). For DS1DH-PBEsol and 1DH-PBEsol, we use $\l=0.80$ (corresponding to a LMP2 fraction of $\l^2=0.64$) which according to Fig.~\ref{MAELE} yields a similar or slightly lower MAE than LMP2. We also report results obtained with the two-parameter DH approximations B2-PLYP, B2GP-PLYP, and mPW2-PLYP, which have smaller fractions of LMP2 ($a_c=0.27$, $0.36$, and $0.25$, respectively). 

Among the non-DH methods, HF, BLYP, B3LYP, and, to a lesser extent, PBE and PBE0, strongly underestimate the lattice energies of the four crystals. PBEsol gives good lattice energies for ammonia, but still underestimated lattice energies for carbon dioxide, and, to a lesser extent, for urea and formamide. This is most likely due to dispersion interactions that are not properly accounted for. 
In fact, the addition of the D3 empirical dispersion correction to PBE and B3LYP leads to much improved results.

A quite uniform accuracy on the lattice energies is also obtained with LMP2. All the DH methods give smaller MAEs that the corresponding Kohn-Sham calculations with the same functionals. However, B2-PLYP, B2GP-PLYP, mPW2-PLYP, DS1DH-BLYP, 1DH-BLYP, DS1DH-PBE, and 1DH-PBE still tend to significantly underestimate the lattice energies. The DS1DH-PBEsol and 1DH-PBEsol approximations give overall reasonably good lattice energies, with a similar average accuracy as LMP2. We note that, whereas most DH methods give smaller lattice energies compared to LMP2, for urea, formamide and ammonia, 1DH-PBEsol gives larger lattice energies than LMP2 and which are closer to the reference values. However, both DS1DH-PBEsol and 1DH-PBEsol give a lattice energy of the carbon dioxide crystal that is significantly more underestimated than in LMP2, suggesting that these DH approximations still miss a part of the dispersion interactions.

\subsection{Comparison of the methods for molecular dimer calculations}
\label{sec:dimers}

\begin{figure*}
\begin{center}
\includegraphics[scale=0.3,angle=-90]{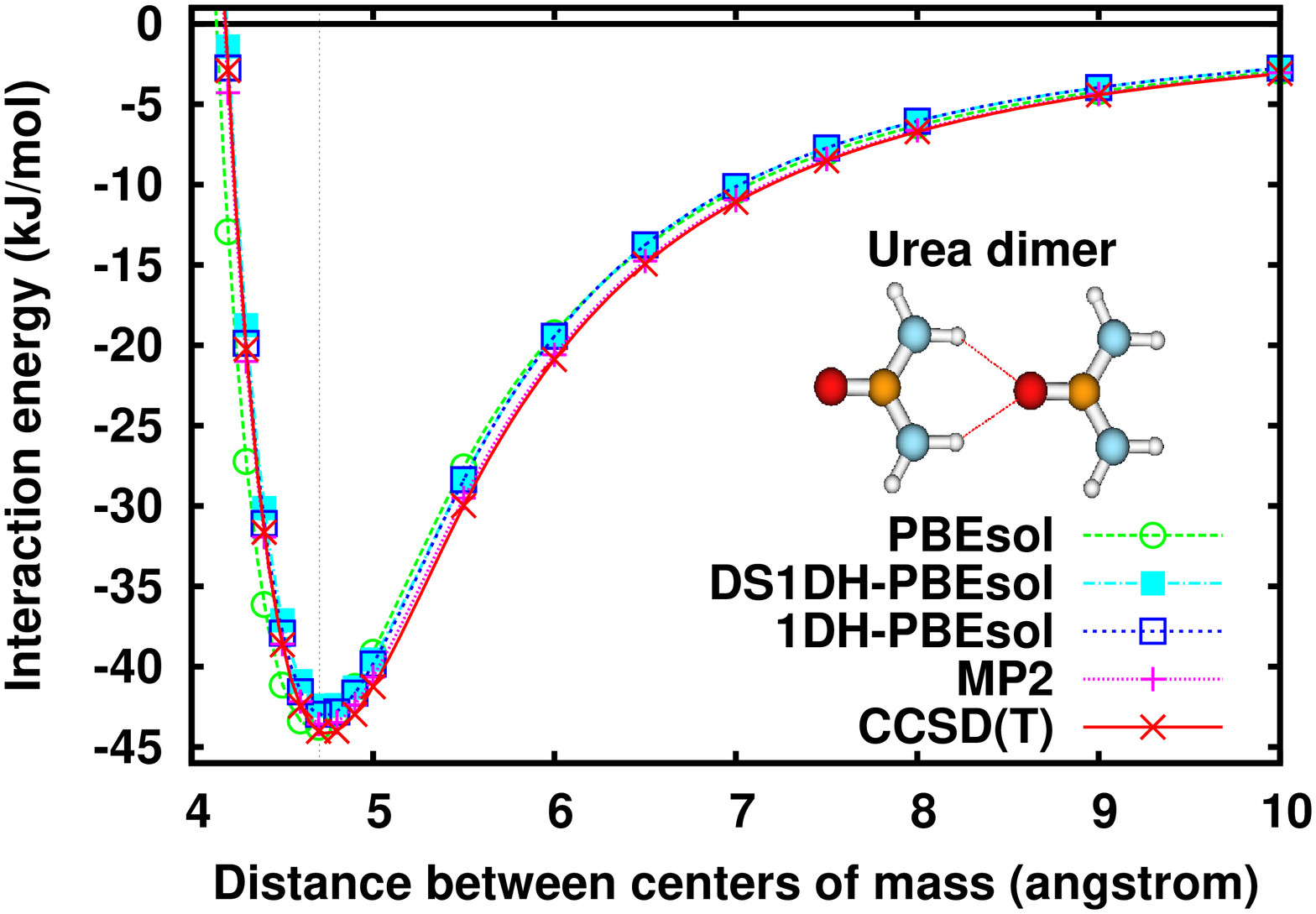}
\includegraphics[scale=0.3,angle=-90]{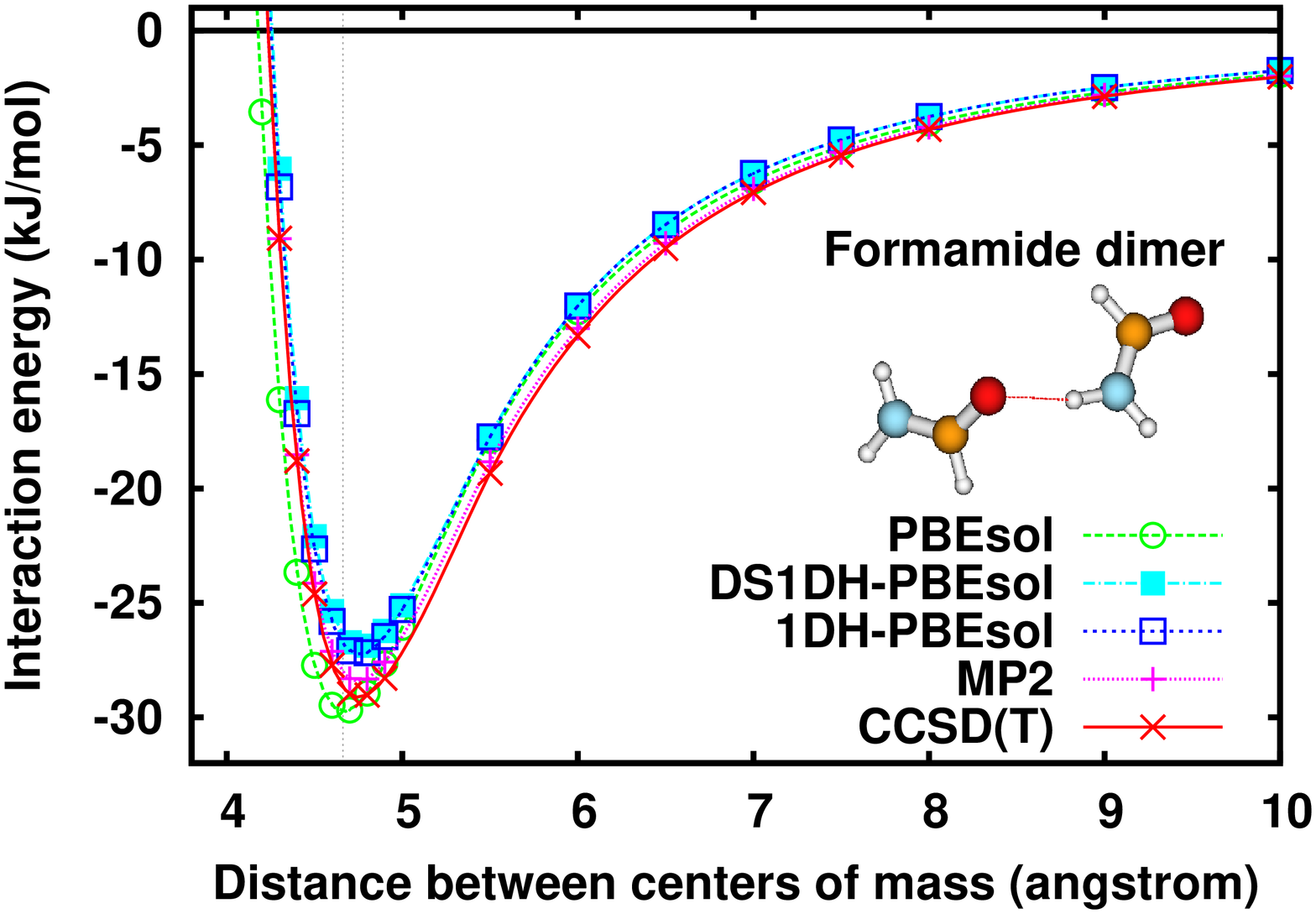}
\includegraphics[scale=0.3,angle=-90]{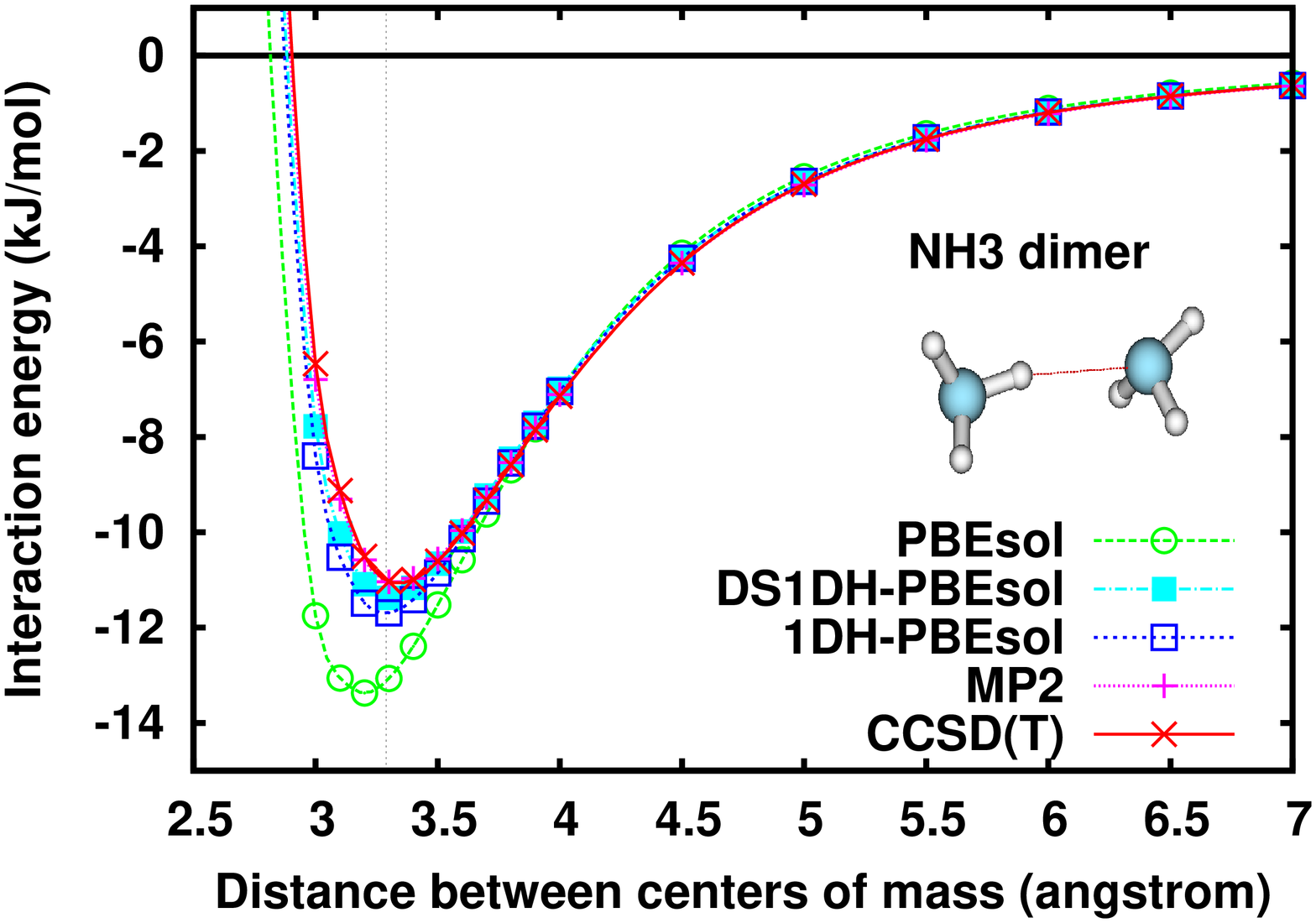}
\includegraphics[scale=0.3,angle=-90]{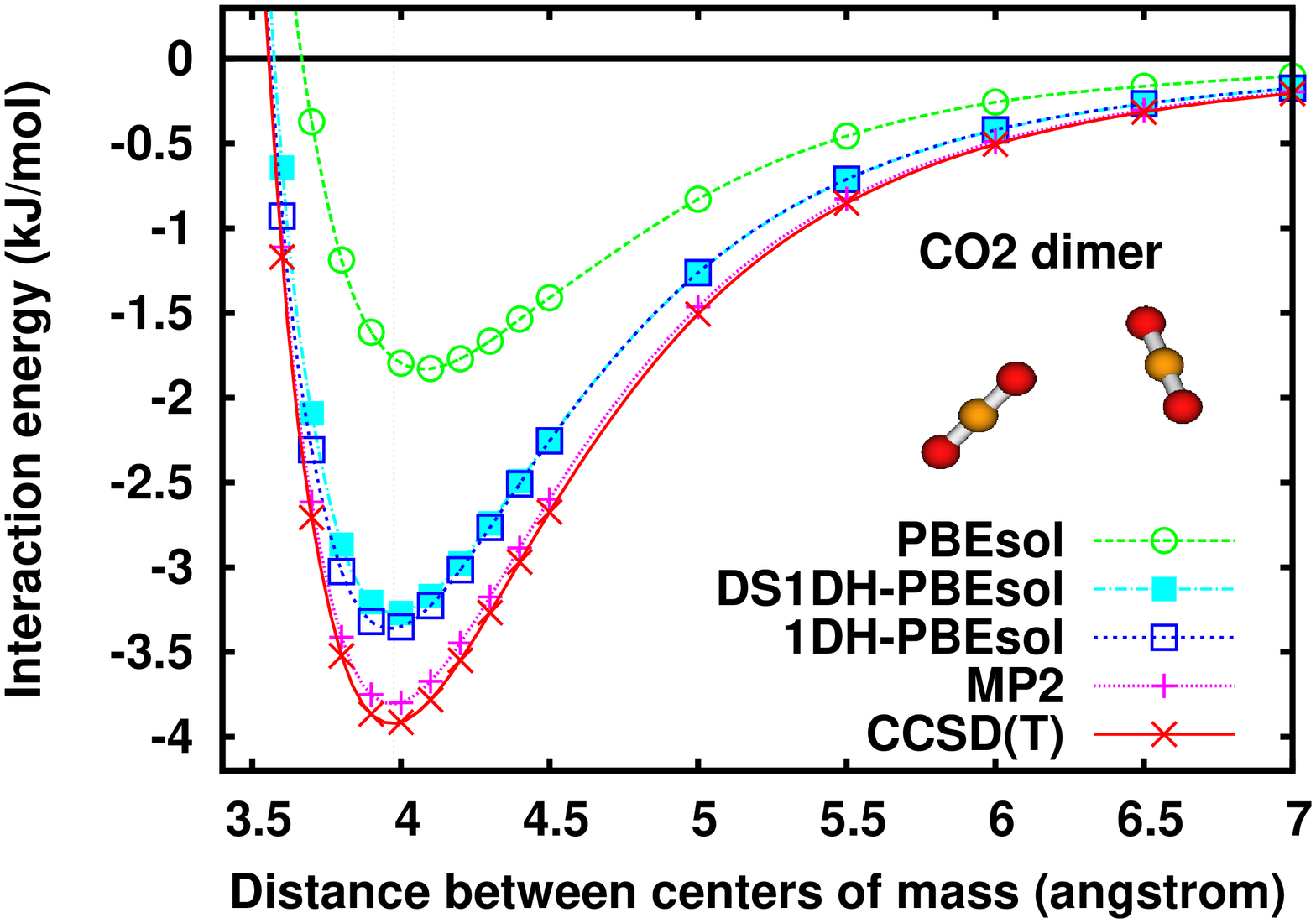}
\end{center}
\caption{Counterpoise-corrected interaction energies (in kJ/mol) of the molecular dimers of urea, formamide, ammonia and carbon dioxide as a function of the distance (in angstrom) between the centers of mass of the molecules, calculated by PBEsol, DS1DH-PBEsol and 1DH-PBEsol ($\l=0.80$), MP2 and CCSD(T), using the aug-cc-pVTZ basis set. The orientation of the molecules in each dimer is fixed to the one of the nearest neighbours in the experimental crystal structure. The vertical dotted line corresponds to the distance in the experimental crystal structure. For urea, the CCSD(T) interaction energy with aug-cc-pVTZ (aVTZ) basis is estimated from the CCSD(T) interaction energy with aug-cc-pVDZ (aVDZ) basis by E$_\text{int}$(CCSD(T)/aVTZ) = E$_\text{int}$(CCSD(T)/aVDZ) + E$_\text{int}$(MP2/aVTZ) - E$_\text{int}$(MP2/aVDZ).}
\label{dimers}
\end{figure*}

In Fig.~\ref{dimers}, we show interaction energy curves of the molecular dimers of urea, formamide, ammonia and carbon dioxide, calculated by PBEsol, DS1DH-PBEsol and 1DH-PBEsol (with $\l=0.80$), MP2 and CCSD(T), using the aug-cc-pVTZ basis set. These dimer calculations allow us to compare the methods with a larger basis set and to have a reliable theoretical reference with CCSD(T). They also permit us to assess the methods for a range of distances between the molecules and to estimate what the effect of the optimization of the lattice parameters in the crystal would be for each method. 

For these dimers, MP2 and CCSD(T) give very close interaction energies for all distances and can be considered as the reference. PBEsol gives quite accurate interaction energies for urea and formamide, but an overestimated interaction energy for ammonia and a largely underestimated interaction energy for carbon dioxide. These deficiencies of PBEsol are corrected by DS1DH-PBEsol and 1DH-PBEsol which give reasonable interaction energy curves for all the dimers, although the interaction energy of the carbon dioxide dimer remains a bit underestimated for all distances. Like in the crystals, neglecting the scaling of the density in the correlation functional, {\it i.e.} going from DS1DH-PBEsol to 1DH-PBEsol, does not lead to significant changes in the interaction energies. For each dimer, the equilibrium intermolecular distance found for each method is rather close to the distance in the experimental crystal structure, so that it unlikely that the optimization of the lattice parameters could change significantly the results. Overall, the calculations on the dimers are consistent with the results obtained for the crystals, and thus give support to the conclusions drawn for crystals.

\subsection{A further benchmark: The hydrogen cyanide crystal}
\label{sec:hcn}

\begin{table*}[t] \centering
\caption{Counterpoise-corrected lattice energies per molecule in (kJ/mol) of the hydrogen cyanide crystal calculated with several methods and basis sets.
}
\begin{tabular}{lrrrr}
\hline
\hline
                              & cc-pVDZ   & cc-pVTZ  &  \phantom{xx} p-aug-cc-pVDZ &  \phantom{xx} p-aug-cc-pVTZ \\
\hline
HF                            &  -10.17   & -17.22   &                &                \\
PBEsol                        &  -32.90   & -34.01   &                &                \\
LMP2                          &  -31.88   & -39.41   &  -41.18$^a$    &    -45.13$^{a,b}$   \\
DS1DH-PBEsol ($\lambda$=0.80) &  -33.57   & -38.84   &  -39.27$^a$    &    -42.00$^a$      \\
1DH-PBEsol ($\lambda$=0.80)   &  -32.69   & -38.02   &  -40.14$^a$    &    -42.80$^a$      \\
\hline
Best theoretical estimate$^c$ &  \multicolumn{4}{r}{-43.5}\\
Experimental values$^d$       &  \multicolumn{4}{r}{-38.7 to -40.5}\\
\hline
\hline
\multicolumn{5}{l}{$^a$With dual-basis set technique: basis sets without augmentation for SCF.}\\
\multicolumn{5}{l}{$^b$The discrepancy between this value and the value of -46.14 of Ref.~\onlinecite{MUsv} is due to}\\[-0.05cm]
\multicolumn{5}{l}{a different computational setup, a slightly different geometry, and the neglect}\\[-0.05cm]
\multicolumn{5}{l}{of molecular relaxation.}\\
\multicolumn{5}{l}{$^c$Ref.~\onlinecite{MUsv}}\\
\multicolumn{5}{l}{$^d$Ref.~\onlinecite{ChiAcr-JPCRD-02}.}\\
\end{tabular}
\label{tab:hcn}
\end{table*}

In order to further assess the performance of the DS1DH-PBEsol and 1DH-PBEsol double hybrids, we now consider the case of the hydrogen cyanide (HCN) crystal. 
Recently, M\"{u}ller and Usvyat~\cite{MUsv} have obtained a coupled-cluster level estimate of the lattice energy per molecule of this crystal, combining periodic LMP2 results and incremental molecular correlation corrections~\cite{Paulus:2006p14134}. Their value, -43.5 kJ/mol, agrees with the experimental values, -38.7 to -40.5 kJ/mol~\cite{ChiAcr-JPCRD-02}, within the experimental uncertainty, while periodic LMP2 in this case overbinds. It is then interesting to study this crystal in our context because, as opposed to systems reported in previous sections, periodic LMP2 method overestimates the interaction energy. Moreover its simple structure (only one molecule -- three atoms -- per unit cell) allows us to use larger correlation-consistent basis sets. The crystal geometry is the same as the one used in Ref.~\onlinecite{MasUsvCiv-CEC-10} and molecular relaxation effects (estimated at around 0.3 kJ/mol)~\cite{MUsv} are neglected. All other computational parameters are consistent with those described in Section~\ref{sec:details}.

The results obtained with HF, PBEsol, LMP2, DS1DH-PBEsol and 1DH-PBEsol ($\lambda$=0.80) using several basis sets are reported in Table~\ref{tab:hcn}. We first notice that, as expected from the high value of $\lambda$, the basis set dependence is only slightly weaker for the double hybrids in comparison to LMP2. The role of augmentation by diffuse functions is once again very important. With the p-aug-cc-pVTZ basis set, DS1DH-PBEsol and 1DH-PBEsol give very similar lattice energies per molecule, -42.0 and -42.8 kJ/mol respectively, which are in good agreement with the best theoretical estimate of -43.5 kJ/mol, and in better agreement with the experimental values than LMP2 is.

\section{Conclusions}
\label{sec:conclusion}

We have implemented a number of double-hybrid approximations in the CRYSTAL09 and CRYSCOR09 suite of programs, and tested them for calculating lattice energies of four molecular crystals: urea, formamide, ammonia, and carbon dioxide. The one-parameter double-hybrid approximations based on the PBEsol density functional, DS1DH-PBEsol and 1DH-PBEsol, with a fraction of HF exchange of $\l=0.80$ and a fraction of LMP2 correlation of $\l^2=0.64$, gives lattice energies per molecule with an accuracy similar to that of LMP2. This conclusion has been further verified on molecular dimers and on the hydrogen cyanide crystal. 

Thus, the present results do not show any clear advantage of the double-hybrid approximations over standard MP2 for molecular crystals since their accuracies are similar as well as their basis-size dependences. Even though this conclusion should be checked on more systems and more properties, we believe that range-separated double-hybrid methods~\cite{AngGerSavTou-PRA-05} with their weak basis-size dependence hold more promise for improving MP2 calculations on molecular crystals. Work is in progress to validate this hypothesis.

\section*{Acknowledgments}

We are grateful to Andreas Savin (Paris), Roberto Dovesi (Torino), and Peter Reinhardt (Paris) for their help with this project.



\begin{thebibliography}{0}
\expandafter\ifx\csname natexlab\endcsname\relax\def\natexlab#1{#1}\fi
\expandafter\ifx\csname bibnamefont\endcsname\relax
  \def\bibnamefont#1{#1}\fi
\expandafter\ifx\csname bibfnamefont\endcsname\relax
  \def\bibfnamefont#1{#1}\fi
\expandafter\ifx\csname citenamefont\endcsname\relax
  \def\citenamefont#1{#1}\fi
\expandafter\ifx\csname url\endcsname\relax
  \def\url#1{\texttt{#1}}\fi
\expandafter\ifx\csname urlprefix\endcsname\relax\def\urlprefix{URL }\fi
\providecommand{\bibinfo}[2]{#2}
\providecommand{\eprint}[2][]{\url{#2}}

\end{thebibliography}


\begin{thebibliography}{106}
\expandafter\ifx\csname natexlab\endcsname\relax\def\natexlab#1{#1}\fi
\expandafter\ifx\csname bibnamefont\endcsname\relax
  \def\bibnamefont#1{#1}\fi
\expandafter\ifx\csname bibfnamefont\endcsname\relax
  \def\bibfnamefont#1{#1}\fi
\expandafter\ifx\csname citenamefont\endcsname\relax
  \def\citenamefont#1{#1}\fi
\expandafter\ifx\csname url\endcsname\relax
  \def\url#1{\texttt{#1}}\fi
\expandafter\ifx\csname urlprefix\endcsname\relax\def\urlprefix{URL }\fi
\providecommand{\bibinfo}[2]{#2}
\providecommand{\eprint}[2][]{\url{#2}}

\bibitem[{\citenamefont{Aaker{\o}y and Seddon}(1993)}]{AakSed-CSR-93}
\bibinfo{author}{\bibfnamefont{C.~B.} \bibnamefont{Aaker{\o}y}}
  \bibnamefont{and} \bibinfo{author}{\bibfnamefont{K.~R.}
  \bibnamefont{Seddon}}, \bibinfo{journal}{Chem. Soc. Rev.}
  \textbf{\bibinfo{volume}{{22}}}, \bibinfo{pages}{397} (\bibinfo{year}{1993}).

\bibitem[{\citenamefont{Karamertzanis et~al.}(2007)\citenamefont{Karamertzanis,
  Anandamanoharan, Fernandes, Cains, Vickers, Tocher, Florence, and
  Price}}]{KarAnaFerCaiVicTocFloPri-JPC-07}
\bibinfo{author}{\bibfnamefont{P.~G.} \bibnamefont{Karamertzanis}},
  \bibinfo{author}{\bibfnamefont{P.~R.} \bibnamefont{Anandamanoharan}},
  \bibinfo{author}{\bibfnamefont{P.}~\bibnamefont{Fernandes}},
  \bibinfo{author}{\bibfnamefont{P.~W.} \bibnamefont{Cains}},
  \bibinfo{author}{\bibfnamefont{M.}~\bibnamefont{Vickers}},
  \bibinfo{author}{\bibfnamefont{D.~A.} \bibnamefont{Tocher}},
  \bibinfo{author}{\bibfnamefont{A.~J.} \bibnamefont{Florence}},
  \bibnamefont{and} \bibinfo{author}{\bibfnamefont{S.~L.} \bibnamefont{Price}},
  \bibinfo{journal}{J. Phys. Chem. B} \textbf{\bibinfo{volume}{111}},
  \bibinfo{pages}{5326} (\bibinfo{year}{2007}).

\bibitem[{\citenamefont{Gourlay et~al.}(2007)\citenamefont{Gourlay, Kendrick,
  and Leusen}}]{GouKenLeu-CGD-07}
\bibinfo{author}{\bibfnamefont{M.~D.} \bibnamefont{Gourlay}},
  \bibinfo{author}{\bibfnamefont{J.}~\bibnamefont{Kendrick}}, \bibnamefont{and}
  \bibinfo{author}{\bibfnamefont{F.~J.~J.} \bibnamefont{Leusen}},
  \bibinfo{journal}{Crystal Growth \& Design} \textbf{\bibinfo{volume}{7}},
  \bibinfo{pages}{56} (\bibinfo{year}{2007}).

\bibitem[{\citenamefont{Kwon et~al.}(2006)\citenamefont{Kwon, Ruiz, Choubey,
  Mutter, Schneider, Jazbinsek, Gramlich, and
  G\"unter}}]{KwoRuiChoMutSchJazGraGun-CM-06}
\bibinfo{author}{\bibfnamefont{O.-P.} \bibnamefont{Kwon}},
  \bibinfo{author}{\bibfnamefont{B.}~\bibnamefont{Ruiz}},
  \bibinfo{author}{\bibfnamefont{A.}~\bibnamefont{Choubey}},
  \bibinfo{author}{\bibfnamefont{L.}~\bibnamefont{Mutter}},
  \bibinfo{author}{\bibfnamefont{A.}~\bibnamefont{Schneider}},
  \bibinfo{author}{\bibfnamefont{M.}~\bibnamefont{Jazbinsek}},
  \bibinfo{author}{\bibfnamefont{V.}~\bibnamefont{Gramlich}}, \bibnamefont{and}
  \bibinfo{author}{\bibfnamefont{P.}~\bibnamefont{G\"unter}},
  \bibinfo{journal}{Chem. Mater.} \textbf{\bibinfo{volume}{18}},
  \bibinfo{pages}{4049} (\bibinfo{year}{2006}).

\bibitem[{\citenamefont{Coombes et~al.}(1996)\citenamefont{Coombes, Price,
  Willock, and Leslie}}]{CooPriWilLes-JPC-96}
\bibinfo{author}{\bibfnamefont{D.~S.} \bibnamefont{Coombes}},
  \bibinfo{author}{\bibfnamefont{S.~L.} \bibnamefont{Price}},
  \bibinfo{author}{\bibfnamefont{D.~J.} \bibnamefont{Willock}},
  \bibnamefont{and} \bibinfo{author}{\bibfnamefont{M.}~\bibnamefont{Leslie}},
  \bibinfo{journal}{J. Phys. Chem.} \textbf{\bibinfo{volume}{100}},
  \bibinfo{pages}{7352} (\bibinfo{year}{1996}).

\bibitem[{\citenamefont{Ferenczy et~al.}(1998)\citenamefont{Ferenczy, Csonka,
  N\'aray-szab\'o, and \'Angy\'an}}]{FerCsoNarAng-JCC-98}
\bibinfo{author}{\bibfnamefont{G.~G.} \bibnamefont{Ferenczy}},
  \bibinfo{author}{\bibfnamefont{G.~I.} \bibnamefont{Csonka}},
  \bibinfo{author}{\bibfnamefont{G.}~\bibnamefont{N\'aray-szab\'o}},
  \bibnamefont{and} \bibinfo{author}{\bibfnamefont{J.~G.}
  \bibnamefont{\'Angy\'an}}, \bibinfo{journal}{J. Comput. Chem.}
  \textbf{\bibinfo{volume}{19}}, \bibinfo{pages}{38} (\bibinfo{year}{1998}).

\bibitem[{\citenamefont{Welch et~al.}(2008)\citenamefont{Welch, Karamertzanis,
  Misquitta, Stone, and Price}}]{WelKarMisStoPri-JCTC-08}
\bibinfo{author}{\bibfnamefont{G.~W.~A.} \bibnamefont{Welch}},
  \bibinfo{author}{\bibfnamefont{P.~G.} \bibnamefont{Karamertzanis}},
  \bibinfo{author}{\bibfnamefont{A.~J.} \bibnamefont{Misquitta}},
  \bibinfo{author}{\bibfnamefont{A.~J.} \bibnamefont{Stone}}, \bibnamefont{and}
  \bibinfo{author}{\bibfnamefont{S.~L.} \bibnamefont{Price}},
  \bibinfo{journal}{J. Chem. Theory Comput.} \textbf{\bibinfo{volume}{4}},
  \bibinfo{pages}{522} (\bibinfo{year}{2008}).

\bibitem[{\citenamefont{Stone and Tsuzuki}(1997)}]{StoTsu-JPCB-97}
\bibinfo{author}{\bibfnamefont{A.~J.} \bibnamefont{Stone}} \bibnamefont{and}
  \bibinfo{author}{\bibfnamefont{S.}~\bibnamefont{Tsuzuki}},
  \bibinfo{journal}{J. Phys. Chem. B} \textbf{\bibinfo{volume}{101}},
  \bibinfo{pages}{10178} (\bibinfo{year}{1997}).

\bibitem[{\citenamefont{Hohenberg and Kohn}(1964)}]{HohKoh-PR-64}
\bibinfo{author}{\bibfnamefont{P.}~\bibnamefont{Hohenberg}} \bibnamefont{and}
  \bibinfo{author}{\bibfnamefont{W.}~\bibnamefont{Kohn}},
  \bibinfo{journal}{Phys. Rev.} \textbf{\bibinfo{volume}{{136}}},
  \bibinfo{pages}{B 864} (\bibinfo{year}{1964}).

\bibitem[{\citenamefont{Kohn and Sham}(1965)}]{KohSha-PR-65}
\bibinfo{author}{\bibfnamefont{W.}~\bibnamefont{Kohn}} \bibnamefont{and}
  \bibinfo{author}{\bibfnamefont{L.~J.} \bibnamefont{Sham}},
  \bibinfo{journal}{Phys. Rev.} \textbf{\bibinfo{volume}{140}},
  \bibinfo{pages}{A1133} (\bibinfo{year}{1965}).

\bibitem[{\citenamefont{Morrison and Siddick}(2003)}]{MorSid-CEJ-03}
\bibinfo{author}{\bibfnamefont{C.~A.} \bibnamefont{Morrison}} \bibnamefont{and}
  \bibinfo{author}{\bibfnamefont{M.~M.} \bibnamefont{Siddick}},
  \bibinfo{journal}{Chem.—Eur. J.} \textbf{\bibinfo{volume}{9}},
  \bibinfo{pages}{628} (\bibinfo{year}{2003}).

\bibitem[{\citenamefont{Fortes et~al.}(2003)\citenamefont{Fortes, Brodholt,
  Wood, and Vocadlo}}]{ForBroWooVoc-JCP-03}
\bibinfo{author}{\bibfnamefont{A.~D.} \bibnamefont{Fortes}},
  \bibinfo{author}{\bibfnamefont{J.~P.} \bibnamefont{Brodholt}},
  \bibinfo{author}{\bibfnamefont{I.~G.} \bibnamefont{Wood}}, \bibnamefont{and}
  \bibinfo{author}{\bibfnamefont{L.}~\bibnamefont{Vocadlo}},
  \bibinfo{journal}{J. Chem. Phys.} \textbf{\bibinfo{volume}{118}},
  \bibinfo{pages}{5987} (\bibinfo{year}{2003}).

\bibitem[{\citenamefont{Ju et~al.}(2005)\citenamefont{Ju, Xiao, and
  Chen}}]{JuXiaCh-IJQC-05}
\bibinfo{author}{\bibfnamefont{X.-H.} \bibnamefont{Ju}},
  \bibinfo{author}{\bibfnamefont{H.-M.} \bibnamefont{Xiao}}, \bibnamefont{and}
  \bibinfo{author}{\bibfnamefont{L.-T.} \bibnamefont{Chen}},
  \bibinfo{journal}{Int. J. Quantum Chem.} \textbf{\bibinfo{volume}{102}},
  \bibinfo{pages}{224} (\bibinfo{year}{2005}).

\bibitem[{\citenamefont{Kristyan and Pulay}(1994)}]{KriPul-CPL-94}
\bibinfo{author}{\bibfnamefont{S.}~\bibnamefont{Kristyan}} \bibnamefont{and}
  \bibinfo{author}{\bibfnamefont{P.}~\bibnamefont{Pulay}},
  \bibinfo{journal}{Chem. Phys. Lett.} \textbf{\bibinfo{volume}{{229}}},
  \bibinfo{pages}{175} (\bibinfo{year}{1994}).

\bibitem[{\citenamefont{Hobza et~al.}(1995)\citenamefont{Hobza, Sponer, and
  Reschel}}]{HobSpoRes-JCC-95}
\bibinfo{author}{\bibfnamefont{P.}~\bibnamefont{Hobza}},
  \bibinfo{author}{\bibfnamefont{J.}~\bibnamefont{Sponer}}, \bibnamefont{and}
  \bibinfo{author}{\bibfnamefont{T.}~\bibnamefont{Reschel}},
  \bibinfo{journal}{J. Comput. Chem.} \textbf{\bibinfo{volume}{16}},
  \bibinfo{pages}{1315} (\bibinfo{year}{1995}).

\bibitem[{\citenamefont{Hongo et~al.}(2010)\citenamefont{Hongo, Watson,
  S\'anchez-Carrera, Iitaka, and Aspuru-Guzik}}]{HonWatSanIitAsp-JPCL-10}
\bibinfo{author}{\bibfnamefont{K.}~\bibnamefont{Hongo}},
  \bibinfo{author}{\bibfnamefont{M.~A.} \bibnamefont{Watson}},
  \bibinfo{author}{\bibfnamefont{R.~S.} \bibnamefont{S\'anchez-Carrera}},
  \bibinfo{author}{\bibfnamefont{T.}~\bibnamefont{Iitaka}}, \bibnamefont{and}
  \bibinfo{author}{\bibfnamefont{A.}~\bibnamefont{Aspuru-Guzik}},
  \bibinfo{journal}{J. Phys. Chem. Lett.} \textbf{\bibinfo{volume}{1}},
  \bibinfo{pages}{1789} (\bibinfo{year}{2010}).

\bibitem[{\citenamefont{Zhao and Truhlar}(2007)}]{ZhaTru-JCTC-07}
\bibinfo{author}{\bibfnamefont{Y.}~\bibnamefont{Zhao}} \bibnamefont{and}
  \bibinfo{author}{\bibfnamefont{D.~G.} \bibnamefont{Truhlar}},
  \bibinfo{journal}{J. Chem. Theory Comput.} \textbf{\bibinfo{volume}{3}},
  \bibinfo{pages}{289} (\bibinfo{year}{2007}).

\bibitem[{\citenamefont{Tsuzuki et~al.}(2010)\citenamefont{Tsuzuki, Orita,
  Honda, and Mikami}}]{TsuOriHonMik-JPCB-10}
\bibinfo{author}{\bibfnamefont{S.}~\bibnamefont{Tsuzuki}},
  \bibinfo{author}{\bibfnamefont{H.}~\bibnamefont{Orita}},
  \bibinfo{author}{\bibfnamefont{K.}~\bibnamefont{Honda}}, \bibnamefont{and}
  \bibinfo{author}{\bibfnamefont{M.}~\bibnamefont{Mikami}},
  \bibinfo{journal}{J. Phys. Chem. B} \textbf{\bibinfo{volume}{114}},
  \bibinfo{pages}{6799} (\bibinfo{year}{2010}).

\bibitem[{\citenamefont{Neumann and Perrin}(2005)}]{NeuPer-JPCB-05}
\bibinfo{author}{\bibfnamefont{M.~A.} \bibnamefont{Neumann}} \bibnamefont{and}
  \bibinfo{author}{\bibfnamefont{M.-A.} \bibnamefont{Perrin}},
  \bibinfo{journal}{J. Phys. Chem. B} \textbf{\bibinfo{volume}{109}},
  \bibinfo{pages}{15531} (\bibinfo{year}{2005}).

\bibitem[{\citenamefont{Li and Feng}(2006)}]{LiFen-PR-06}
\bibinfo{author}{\bibfnamefont{T.}~\bibnamefont{Li}} \bibnamefont{and}
  \bibinfo{author}{\bibfnamefont{S.}~\bibnamefont{Feng}},
  \bibinfo{journal}{Pharm. Res.} \textbf{\bibinfo{volume}{23}},
  \bibinfo{pages}{2326} (\bibinfo{year}{2006}).

\bibitem[{\citenamefont{Civalleri et~al.}(2008)\citenamefont{Civalleri,
  Zicovich-Wilson, Valenzano, and Ugliengo}}]{Civalleri:2008p37490}
\bibinfo{author}{\bibfnamefont{B.}~\bibnamefont{Civalleri}},
  \bibinfo{author}{\bibfnamefont{C.}~\bibnamefont{Zicovich-Wilson}},
  \bibinfo{author}{\bibfnamefont{L.}~\bibnamefont{Valenzano}},
  \bibnamefont{and} \bibinfo{author}{\bibfnamefont{P.}~\bibnamefont{Ugliengo}},
  \bibinfo{journal}{Cryst. Eng. Comm} \textbf{\bibinfo{volume}{10}},
  \bibinfo{pages}{405} (\bibinfo{year}{2008}).

\bibitem[{\citenamefont{Karamertzanis et~al.}(2008)\citenamefont{Karamertzanis,
  Day, Welch, Kendrick, Leusen, Neumann, and
  Price}}]{KarDayWelKenLeuNeuPri-JPC-08}
\bibinfo{author}{\bibfnamefont{P.~G.} \bibnamefont{Karamertzanis}},
  \bibinfo{author}{\bibfnamefont{G.~M.} \bibnamefont{Day}},
  \bibinfo{author}{\bibfnamefont{G.~W.~A.} \bibnamefont{Welch}},
  \bibinfo{author}{\bibfnamefont{J.}~\bibnamefont{Kendrick}},
  \bibinfo{author}{\bibfnamefont{F.~J.~J.} \bibnamefont{Leusen}},
  \bibinfo{author}{\bibfnamefont{M.~A.} \bibnamefont{Neumann}},
  \bibnamefont{and} \bibinfo{author}{\bibfnamefont{S.~L.} \bibnamefont{Price}},
  \bibinfo{journal}{J. Phys. Chem.} \textbf{\bibinfo{volume}{128}},
  \bibinfo{pages}{244708} (\bibinfo{year}{2008}).

\bibitem[{\citenamefont{Ugliengo et~al.}(2009)\citenamefont{Ugliengo,
  Zicovich-Wilson, Tosoni, and Civalleri}}]{UglZicTosCiv-JMC-09}
\bibinfo{author}{\bibfnamefont{P.}~\bibnamefont{Ugliengo}},
  \bibinfo{author}{\bibfnamefont{C.}~\bibnamefont{Zicovich-Wilson}},
  \bibinfo{author}{\bibfnamefont{S.}~\bibnamefont{Tosoni}}, \bibnamefont{and}
  \bibinfo{author}{\bibfnamefont{B.}~\bibnamefont{Civalleri}},
  \bibinfo{journal}{J. Mater. Chem.} \textbf{\bibinfo{volume}{19}},
  \bibinfo{pages}{2564} (\bibinfo{year}{2009}).

\bibitem[{\citenamefont{Zicovich-Wilson
  et~al.}(2010)\citenamefont{Zicovich-Wilson, Kirtman, Civalleri, and
  Ram\'irez-Sol\'is}}]{ZicKirCivRam-PCCP-10}
\bibinfo{author}{\bibfnamefont{C.~M.} \bibnamefont{Zicovich-Wilson}},
  \bibinfo{author}{\bibfnamefont{B.}~\bibnamefont{Kirtman}},
  \bibinfo{author}{\bibfnamefont{B.}~\bibnamefont{Civalleri}},
  \bibnamefont{and}
  \bibinfo{author}{\bibfnamefont{A.}~\bibnamefont{Ram\'irez-Sol\'is}},
  \bibinfo{journal}{Phys. Chem. Chem. Phys.} \textbf{\bibinfo{volume}{12}},
  \bibinfo{pages}{3289} (\bibinfo{year}{2010}).

\bibitem[{\citenamefont{Sorescu and Rice}(2010)}]{SorRic-JPCC-10}
\bibinfo{author}{\bibfnamefont{D.~C.} \bibnamefont{Sorescu}} \bibnamefont{and}
  \bibinfo{author}{\bibfnamefont{B.~M.} \bibnamefont{Rice}},
  \bibinfo{journal}{J. Phys. Chem. C} \textbf{\bibinfo{volume}{114}},
  \bibinfo{pages}{6734} (\bibinfo{year}{2010}).

\bibitem[{\citenamefont{Balu et~al.}(2011)\citenamefont{Balu, Byrd, and
  Rice}}]{BalByrRic-JPCB-11}
\bibinfo{author}{\bibfnamefont{R.}~\bibnamefont{Balu}},
  \bibinfo{author}{\bibfnamefont{E.~F.~C.} \bibnamefont{Byrd}},
  \bibnamefont{and} \bibinfo{author}{\bibfnamefont{B.~M.} \bibnamefont{Rice}},
  \bibinfo{journal}{J. Phys. Chem. B} \textbf{\bibinfo{volume}{115}},
  \bibinfo{pages}{803} (\bibinfo{year}{2011}).

\bibitem[{\citenamefont{Reckien et~al.}(2012)\citenamefont{Reckien, Janetzko,
  Peintinger, and Bredow}}]{RecJanPeiBre-JCC-12}
\bibinfo{author}{\bibfnamefont{W.}~\bibnamefont{Reckien}},
  \bibinfo{author}{\bibfnamefont{F.}~\bibnamefont{Janetzko}},
  \bibinfo{author}{\bibfnamefont{M.~F.} \bibnamefont{Peintinger}},
  \bibnamefont{and} \bibinfo{author}{\bibfnamefont{T.}~\bibnamefont{Bredow}},
  \bibinfo{journal}{J. Comput. Chem.} \textbf{\bibinfo{volume}{{33}}},
  \bibinfo{pages}{2023} (\bibinfo{year}{2012}).

\bibitem[{\citenamefont{{A. Otero-de-la-Roza and E. R.
  Johnson}}(2012{\natexlab{a}})}]{OteJoh-JCP-12a}
\bibinfo{author}{\bibnamefont{{A. Otero-de-la-Roza and E. R. Johnson}}},
  \bibinfo{journal}{J. Chem. Phys.} \textbf{\bibinfo{volume}{136}},
  \bibinfo{pages}{174109} (\bibinfo{year}{2012}{\natexlab{a}}).

\bibitem[{\citenamefont{{A. Otero-de-la-Roza and E. R.
  Johnson}}(2012{\natexlab{b}})}]{OteJoh-JCP-12}
\bibinfo{author}{\bibnamefont{{A. Otero-de-la-Roza and E. R. Johnson}}},
  \bibinfo{journal}{J. Chem. Phys.} \textbf{\bibinfo{volume}{137}},
  \bibinfo{pages}{054103} (\bibinfo{year}{2012}{\natexlab{b}}).

\bibitem[{\citenamefont{Dion et~al.}(2004)\citenamefont{Dion, Rydberg,
  Schr\"oder, Langreth, and Lundqvist}}]{DioRydSchLan-PRL-04}
\bibinfo{author}{\bibfnamefont{M.}~\bibnamefont{Dion}},
  \bibinfo{author}{\bibfnamefont{H.}~\bibnamefont{Rydberg}},
  \bibinfo{author}{\bibfnamefont{E.}~\bibnamefont{Schr\"oder}},
  \bibinfo{author}{\bibfnamefont{D.~C.} \bibnamefont{Langreth}},
  \bibnamefont{and} \bibinfo{author}{\bibfnamefont{B.~I.}
  \bibnamefont{Lundqvist}}, \bibinfo{journal}{Phys. Rev. Lett.}
  \textbf{\bibinfo{volume}{92}}, \bibinfo{pages}{246401}
  (\bibinfo{year}{2004}).

\bibitem[{\citenamefont{Thonhauser et~al.}(2007)\citenamefont{Thonhauser,
  Cooper, Li, Puzder, Hyldgaard, and Langreth}}]{ThoCooLiPuzHylLan-PRB-07}
\bibinfo{author}{\bibfnamefont{T.}~\bibnamefont{Thonhauser}},
  \bibinfo{author}{\bibfnamefont{V.~R.} \bibnamefont{Cooper}},
  \bibinfo{author}{\bibfnamefont{S.}~\bibnamefont{Li}},
  \bibinfo{author}{\bibfnamefont{A.}~\bibnamefont{Puzder}},
  \bibinfo{author}{\bibfnamefont{P.}~\bibnamefont{Hyldgaard}},
  \bibnamefont{and} \bibinfo{author}{\bibfnamefont{D.~C.}
  \bibnamefont{Langreth}}, \bibinfo{journal}{Phys. Rev. B.}
  \textbf{\bibinfo{volume}{76}}, \bibinfo{pages}{125112}
  (\bibinfo{year}{2007}).

\bibitem[{\citenamefont{Shimojo et~al.}(2010)\citenamefont{Shimojo, Wu, Nakano,
  Kalia, and Vashishta}}]{ShiWuNakKalVas-JCP-10}
\bibinfo{author}{\bibfnamefont{F.}~\bibnamefont{Shimojo}},
  \bibinfo{author}{\bibfnamefont{Z.}~\bibnamefont{Wu}},
  \bibinfo{author}{\bibfnamefont{A.}~\bibnamefont{Nakano}},
  \bibinfo{author}{\bibfnamefont{R.~K.} \bibnamefont{Kalia}}, \bibnamefont{and}
  \bibinfo{author}{\bibfnamefont{P.}~\bibnamefont{Vashishta}},
  \bibinfo{journal}{J. Chem. Phys.} \textbf{\bibinfo{volume}{132}},
  \bibinfo{pages}{094106} (\bibinfo{year}{2010}).

\bibitem[{\citenamefont{Alfredsson et~al.}(1996)\citenamefont{Alfredsson,
  Ojamae, and Hermansson}}]{AlfOjaHer-IJQC-96}
\bibinfo{author}{\bibfnamefont{M.}~\bibnamefont{Alfredsson}},
  \bibinfo{author}{\bibfnamefont{L.}~\bibnamefont{Ojamae}}, \bibnamefont{and}
  \bibinfo{author}{\bibfnamefont{K.~G.} \bibnamefont{Hermansson}},
  \bibinfo{journal}{Int. J. Quantum Chem.} \textbf{\bibinfo{volume}{60}},
  \bibinfo{pages}{767} (\bibinfo{year}{1996}).

\bibitem[{\citenamefont{Ro\'sciszewski
  et~al.}(1999)\citenamefont{Ro\'sciszewski, Paulus, Fulde, and
  Stoll}}]{RosPauFulSto-PRB-99}
\bibinfo{author}{\bibfnamefont{K.}~\bibnamefont{Ro\'sciszewski}},
  \bibinfo{author}{\bibfnamefont{B.}~\bibnamefont{Paulus}},
  \bibinfo{author}{\bibfnamefont{P.}~\bibnamefont{Fulde}}, \bibnamefont{and}
  \bibinfo{author}{\bibfnamefont{H.}~\bibnamefont{Stoll}},
  \bibinfo{journal}{Phys. Rev. B} \textbf{\bibinfo{volume}{60}},
  \bibinfo{pages}{7905} (\bibinfo{year}{1999}).

\bibitem[{\citenamefont{Ikeda et~al.}(2003)\citenamefont{Ikeda, Nagayoshi, and
  Kitaura}}]{IkeNagKit-CPL-03}
\bibinfo{author}{\bibfnamefont{T.}~\bibnamefont{Ikeda}},
  \bibinfo{author}{\bibfnamefont{K.}~\bibnamefont{Nagayoshi}},
  \bibnamefont{and} \bibinfo{author}{\bibfnamefont{K.}~\bibnamefont{Kitaura}},
  \bibinfo{journal}{Chem. Phys. Lett.} \textbf{\bibinfo{volume}{370}},
  \bibinfo{pages}{218} (\bibinfo{year}{2003}).

\bibitem[{\citenamefont{Ringer and Sherrill}(2008)}]{RinShe-CE-08}
\bibinfo{author}{\bibfnamefont{A.~L.} \bibnamefont{Ringer}} \bibnamefont{and}
  \bibinfo{author}{\bibfnamefont{C.~D.} \bibnamefont{Sherrill}},
  \bibinfo{journal}{Chem. Eur. J.} \textbf{\bibinfo{volume}{14}},
  \bibinfo{pages}{2542} (\bibinfo{year}{2008}).

\bibitem[{\citenamefont{Podeszwa et~al.}(2008)\citenamefont{Podeszwa, Rice, and
  Szalewicz}}]{PodRicSza-PRL-08}
\bibinfo{author}{\bibfnamefont{R.}~\bibnamefont{Podeszwa}},
  \bibinfo{author}{\bibfnamefont{B.~M.} \bibnamefont{Rice}}, \bibnamefont{and}
  \bibinfo{author}{\bibfnamefont{K.}~\bibnamefont{Szalewicz}},
  \bibinfo{journal}{Phys. Rev. Lett.} \textbf{\bibinfo{volume}{101}},
  \bibinfo{pages}{115503} (\bibinfo{year}{2008}).

\bibitem[{\citenamefont{Hermann and Schwerdtfeger}(2008)}]{HerSch-PRL-08}
\bibinfo{author}{\bibfnamefont{A.}~\bibnamefont{Hermann}} \bibnamefont{and}
  \bibinfo{author}{\bibfnamefont{P.}~\bibnamefont{Schwerdtfeger}},
  \bibinfo{journal}{Phys. Rev. Lett.} \textbf{\bibinfo{volume}{101}},
  \bibinfo{pages}{183005} (\bibinfo{year}{2008}).

\bibitem[{\citenamefont{Wen and Beran}(2011)}]{WenBer-JCTC-11}
\bibinfo{author}{\bibfnamefont{S.}~\bibnamefont{Wen}} \bibnamefont{and}
  \bibinfo{author}{\bibfnamefont{G.~J.~O.} \bibnamefont{Beran}},
  \bibinfo{journal}{J. Chem. Theory Comput.} \textbf{\bibinfo{volume}{7}},
  \bibinfo{pages}{3733} (\bibinfo{year}{2011}).

\bibitem[{\citenamefont{Bygrave et~al.}(2012)\citenamefont{Bygrave, Allan, and
  Manby}}]{BygAllMan-JCP-12}
\bibinfo{author}{\bibfnamefont{P.~J.} \bibnamefont{Bygrave}},
  \bibinfo{author}{\bibfnamefont{N.~L.} \bibnamefont{Allan}}, \bibnamefont{and}
  \bibinfo{author}{\bibfnamefont{F.~R.} \bibnamefont{Manby}},
  \bibinfo{journal}{J. Chem. Phys.} \textbf{\bibinfo{volume}{137}},
  \bibinfo{pages}{164102} (\bibinfo{year}{2012}).

\bibitem[{\citenamefont{Sode et~al.}(2013)\citenamefont{Sode, Ke\c{c}eli, Yagi,
  and Hirata}}]{SodKecYagHir-PRB-13}
\bibinfo{author}{\bibfnamefont{O.}~\bibnamefont{Sode}},
  \bibinfo{author}{\bibfnamefont{M.}~\bibnamefont{Ke\c{c}eli}},
  \bibinfo{author}{\bibfnamefont{K.}~\bibnamefont{Yagi}}, \bibnamefont{and}
  \bibinfo{author}{\bibfnamefont{S.}~\bibnamefont{Hirata}},
  \bibinfo{journal}{Phys. Rev. B} \textbf{\bibinfo{volume}{138}},
  \bibinfo{pages}{074501} (\bibinfo{year}{2013}).

\bibitem[{\citenamefont{Sun and Bartlett}(1996)}]{SunBar-JCP-96}
\bibinfo{author}{\bibfnamefont{J.-Q.} \bibnamefont{Sun}} \bibnamefont{and}
  \bibinfo{author}{\bibfnamefont{R.~J.} \bibnamefont{Bartlett}},
  \bibinfo{journal}{J. Chem. Phys.} \textbf{\bibinfo{volume}{104}},
  \bibinfo{pages}{8553} (\bibinfo{year}{1996}).

\bibitem[{\citenamefont{Ayala et~al.}(2001)\citenamefont{Ayala, Kudin, and
  Scuseria}}]{AyaKudScu-JCP-01}
\bibinfo{author}{\bibfnamefont{P.~Y.} \bibnamefont{Ayala}},
  \bibinfo{author}{\bibfnamefont{K.~N.} \bibnamefont{Kudin}}, \bibnamefont{and}
  \bibinfo{author}{\bibfnamefont{G.~E.} \bibnamefont{Scuseria}},
  \bibinfo{journal}{J. Chem. Phys.} \textbf{\bibinfo{volume}{115}},
  \bibinfo{pages}{9698} (\bibinfo{year}{2001}).

\bibitem[{\citenamefont{Marsman et~al.}(2009)\citenamefont{Marsman, Gr\"uneis,
  Paier, and Kresse}}]{MarGruPaiKre-JCP-09}
\bibinfo{author}{\bibfnamefont{M.}~\bibnamefont{Marsman}},
  \bibinfo{author}{\bibfnamefont{A.}~\bibnamefont{Gr\"uneis}},
  \bibinfo{author}{\bibfnamefont{J.}~\bibnamefont{Paier}}, \bibnamefont{and}
  \bibinfo{author}{\bibfnamefont{G.}~\bibnamefont{Kresse}},
  \bibinfo{journal}{J. Chem. Phys.} \textbf{\bibinfo{volume}{130}},
  \bibinfo{pages}{184103} (\bibinfo{year}{2009}).

\bibitem[{\citenamefont{Gr\"uneis et~al.}(2010)\citenamefont{Gr\"uneis,
  Marsman, and Kresse}}]{GruMarKre-JCP-10}
\bibinfo{author}{\bibfnamefont{A.}~\bibnamefont{Gr\"uneis}},
  \bibinfo{author}{\bibfnamefont{M.}~\bibnamefont{Marsman}}, \bibnamefont{and}
  \bibinfo{author}{\bibfnamefont{G.}~\bibnamefont{Kresse}},
  \bibinfo{journal}{J. Chem. Phys.} \textbf{\bibinfo{volume}{133}},
  \bibinfo{pages}{074107} (\bibinfo{year}{2010}).

\bibitem[{\citenamefont{{M. Del Ben, J. Hutter, and J.
  VandeVondele}}(2012)}]{DelHutVan-JCTC-12}
\bibinfo{author}{\bibnamefont{{M. Del Ben, J. Hutter, and J. VandeVondele}}},
  \bibinfo{journal}{J. Chem. Theory Comput.} \textbf{\bibinfo{volume}{{8}}},
  \bibinfo{pages}{4177} (\bibinfo{year}{2012}).

\bibitem[{\citenamefont{{M. Del Ben, J. Hutter, and J.
  VandeVondele}}(2013)}]{DelHutVan-JCTC-13}
\bibinfo{author}{\bibnamefont{{M. Del Ben, J. Hutter, and J. VandeVondele}}},
  \bibinfo{journal}{J. Chem. Theory Comput.} \textbf{\bibinfo{volume}{{9}}},
  \bibinfo{pages}{2654} (\bibinfo{year}{2013}).

\bibitem[{\citenamefont{Pisani et~al.}(2005)\citenamefont{Pisani, Busso,
  Capecchi, Casassa, Dovesi, Maschio, Zicovich-Wilson, and
  Sch\"utz}}]{PisBusCapCasDovMasZicSch-JCP-05}
\bibinfo{author}{\bibfnamefont{C.}~\bibnamefont{Pisani}},
  \bibinfo{author}{\bibfnamefont{M.}~\bibnamefont{Busso}},
  \bibinfo{author}{\bibfnamefont{G.}~\bibnamefont{Capecchi}},
  \bibinfo{author}{\bibfnamefont{S.}~\bibnamefont{Casassa}},
  \bibinfo{author}{\bibfnamefont{R.}~\bibnamefont{Dovesi}},
  \bibinfo{author}{\bibfnamefont{L.}~\bibnamefont{Maschio}},
  \bibinfo{author}{\bibfnamefont{C.}~\bibnamefont{Zicovich-Wilson}},
  \bibnamefont{and} \bibinfo{author}{\bibfnamefont{M.}~\bibnamefont{Sch\"utz}},
  \bibinfo{journal}{J. Chem. Phys.} \textbf{\bibinfo{volume}{{122}}},
  \bibinfo{pages}{094113} (\bibinfo{year}{2005}).

\bibitem[{\citenamefont{Lu et~al.}(2009)\citenamefont{Lu, Li, Rocca, and
  Galli}}]{LuLiRocGal-PRL-09}
\bibinfo{author}{\bibfnamefont{D.}~\bibnamefont{Lu}},
  \bibinfo{author}{\bibfnamefont{Y.}~\bibnamefont{Li}},
  \bibinfo{author}{\bibfnamefont{D.}~\bibnamefont{Rocca}}, \bibnamefont{and}
  \bibinfo{author}{\bibfnamefont{G.}~\bibnamefont{Galli}},
  \bibinfo{journal}{Phys. Rev. Lett.} \textbf{\bibinfo{volume}{102}},
  \bibinfo{pages}{206411} (\bibinfo{year}{2009}).

\bibitem[{\citenamefont{Li et~al.}(2010)\citenamefont{Li, Lu, Nguyen, and
  Galli}}]{LiLuNguGal-JPCA-10}
\bibinfo{author}{\bibfnamefont{Y.}~\bibnamefont{Li}},
  \bibinfo{author}{\bibfnamefont{D.}~\bibnamefont{Lu}},
  \bibinfo{author}{\bibfnamefont{H.-V.} \bibnamefont{Nguyen}},
  \bibnamefont{and} \bibinfo{author}{\bibfnamefont{G.}~\bibnamefont{Galli}},
  \bibinfo{journal}{J. Phys. Chem. A} \textbf{\bibinfo{volume}{114}},
  \bibinfo{pages}{1944} (\bibinfo{year}{2010}).

\bibitem[{\citenamefont{Maschio}(2011)}]{Mas-JCTC-11}
\bibinfo{author}{\bibfnamefont{L.}~\bibnamefont{Maschio}}, \bibinfo{journal}{J.
  Chem. Theory Comput.} \textbf{\bibinfo{volume}{{7}}}, \bibinfo{pages}{2818}
  (\bibinfo{year}{2011}).

\bibitem[{\citenamefont{Pisani et~al.}(2012)\citenamefont{Pisani, Sch\"utz,
  Casassa, Usvyat, Maschio, Lorenz, and Erba}}]{PisSchCasUsvMasLorErb-PCCP-12}
\bibinfo{author}{\bibfnamefont{C.}~\bibnamefont{Pisani}},
  \bibinfo{author}{\bibfnamefont{M.}~\bibnamefont{Sch\"utz}},
  \bibinfo{author}{\bibfnamefont{S.}~\bibnamefont{Casassa}},
  \bibinfo{author}{\bibfnamefont{D.}~\bibnamefont{Usvyat}},
  \bibinfo{author}{\bibfnamefont{L.}~\bibnamefont{Maschio}},
  \bibinfo{author}{\bibfnamefont{M.}~\bibnamefont{Lorenz}}, \bibnamefont{and}
  \bibinfo{author}{\bibfnamefont{A.}~\bibnamefont{Erba}},
  \bibinfo{journal}{Phys. Chem. Chem. Phys.} \textbf{\bibinfo{volume}{{14}}},
  \bibinfo{pages}{7615} (\bibinfo{year}{2012}).

\bibitem[{\citenamefont{Casassa et~al.}(2008)\citenamefont{Casassa, Halo, and
  Maschio}}]{Casassa:2008p54747}
\bibinfo{author}{\bibfnamefont{S.}~\bibnamefont{Casassa}},
  \bibinfo{author}{\bibfnamefont{M.}~\bibnamefont{Halo}}, \bibnamefont{and}
  \bibinfo{author}{\bibfnamefont{L.}~\bibnamefont{Maschio}},
  \bibinfo{journal}{J. Phys.: Conf. Ser} \textbf{\bibinfo{volume}{117}},
  \bibinfo{pages}{012007} (\bibinfo{year}{2008}).

\bibitem[{\citenamefont{Halo et~al.}(2009{\natexlab{a}})\citenamefont{Halo,
  Casassa, Maschio, and Pisani}}]{Halo:2009p30779}
\bibinfo{author}{\bibfnamefont{M.}~\bibnamefont{Halo}},
  \bibinfo{author}{\bibfnamefont{S.}~\bibnamefont{Casassa}},
  \bibinfo{author}{\bibfnamefont{L.}~\bibnamefont{Maschio}}, \bibnamefont{and}
  \bibinfo{author}{\bibfnamefont{C.}~\bibnamefont{Pisani}},
  \bibinfo{journal}{Chem. Phys. Lett.} \textbf{\bibinfo{volume}{467}},
  \bibinfo{pages}{294} (\bibinfo{year}{2009}{\natexlab{a}}).

\bibitem[{\citenamefont{Halo et~al.}(2009{\natexlab{b}})\citenamefont{Halo,
  Casassa, Maschio, and Pisani}}]{Halo:2009p30774}
\bibinfo{author}{\bibfnamefont{M.}~\bibnamefont{Halo}},
  \bibinfo{author}{\bibfnamefont{S.}~\bibnamefont{Casassa}},
  \bibinfo{author}{\bibfnamefont{L.}~\bibnamefont{Maschio}}, \bibnamefont{and}
  \bibinfo{author}{\bibfnamefont{C.}~\bibnamefont{Pisani}},
  \bibinfo{journal}{Phys. Chem. Chem. Phys.} \textbf{\bibinfo{volume}{11}},
  \bibinfo{pages}{586} (\bibinfo{year}{2009}{\natexlab{b}}).

\bibitem[{\citenamefont{Maschio
  et~al.}(2010{\natexlab{a}})\citenamefont{Maschio, Usvyat, Sch\"utz, and
  Civalleri}}]{MasUsvSchCiv-JCP-10}
\bibinfo{author}{\bibfnamefont{L.}~\bibnamefont{Maschio}},
  \bibinfo{author}{\bibfnamefont{D.}~\bibnamefont{Usvyat}},
  \bibinfo{author}{\bibfnamefont{M.}~\bibnamefont{Sch\"utz}}, \bibnamefont{and}
  \bibinfo{author}{\bibfnamefont{B.}~\bibnamefont{Civalleri}},
  \bibinfo{journal}{J. Chem. Phys.} \textbf{\bibinfo{volume}{132}},
  \bibinfo{pages}{134706} (\bibinfo{year}{2010}{\natexlab{a}}).

\bibitem[{\citenamefont{Maschio
  et~al.}(2010{\natexlab{b}})\citenamefont{Maschio, Usvyat, and
  Civalleri}}]{MasUsvCiv-CEC-10}
\bibinfo{author}{\bibfnamefont{L.}~\bibnamefont{Maschio}},
  \bibinfo{author}{\bibfnamefont{D.}~\bibnamefont{Usvyat}}, \bibnamefont{and}
  \bibinfo{author}{\bibfnamefont{B.}~\bibnamefont{Civalleri}},
  \bibinfo{journal}{CrystEngComm} \textbf{\bibinfo{volume}{12}},
  \bibinfo{pages}{2429} (\bibinfo{year}{2010}{\natexlab{b}}).

\bibitem[{\citenamefont{Maschio et~al.}(2011)\citenamefont{Maschio, Civalleri,
  Ugliengo, and Gavezzotti}}]{MasCivUglGav-JPC-11}
\bibinfo{author}{\bibfnamefont{L.}~\bibnamefont{Maschio}},
  \bibinfo{author}{\bibfnamefont{B.}~\bibnamefont{Civalleri}},
  \bibinfo{author}{\bibfnamefont{P.}~\bibnamefont{Ugliengo}}, \bibnamefont{and}
  \bibinfo{author}{\bibfnamefont{A.}~\bibnamefont{Gavezzotti}},
  \bibinfo{journal}{J. Phys. Chem. A} \textbf{\bibinfo{volume}{115}},
  \bibinfo{pages}{11179} (\bibinfo{year}{2011}).

\bibitem[{\citenamefont{Grimme}(2006)}]{Gri-JCP-06}
\bibinfo{author}{\bibfnamefont{S.}~\bibnamefont{Grimme}}, \bibinfo{journal}{J.
  Chem. Phys.} \textbf{\bibinfo{volume}{124}}, \bibinfo{pages}{034108}
  (\bibinfo{year}{2006}).

\bibitem[{\citenamefont{Becke}(1988)}]{Bec-PRA-88}
\bibinfo{author}{\bibfnamefont{A.~D.} \bibnamefont{Becke}},
  \bibinfo{journal}{Phys. Rev. A} \textbf{\bibinfo{volume}{{38}}},
  \bibinfo{pages}{3098} (\bibinfo{year}{1988}).

\bibitem[{\citenamefont{Lee et~al.}(1988)\citenamefont{Lee, Yang, and
  Parr}}]{LeeYanPar-PRB-88}
\bibinfo{author}{\bibfnamefont{C.}~\bibnamefont{Lee}},
  \bibinfo{author}{\bibfnamefont{W.}~\bibnamefont{Yang}}, \bibnamefont{and}
  \bibinfo{author}{\bibfnamefont{R.~G.} \bibnamefont{Parr}},
  \bibinfo{journal}{Phys. Rev. B} \textbf{\bibinfo{volume}{37}},
  \bibinfo{pages}{785} (\bibinfo{year}{1988}).

\bibitem[{\citenamefont{Sancho-Garc\'ia
  et~al.}(2013)\citenamefont{Sancho-Garc\'ia, Arag\'o, Ort\'i, and
  Olivier}}]{SanAraOrtOli-JCP-13}
\bibinfo{author}{\bibfnamefont{J.~C.} \bibnamefont{Sancho-Garc\'ia}},
  \bibinfo{author}{\bibfnamefont{J.}~\bibnamefont{Arag\'o}},
  \bibinfo{author}{\bibfnamefont{E.}~\bibnamefont{Ort\'i}}, \bibnamefont{and}
  \bibinfo{author}{\bibfnamefont{Y.}~\bibnamefont{Olivier}},
  \bibinfo{journal}{J. Chem. Phys.} \textbf{\bibinfo{volume}{138}},
  \bibinfo{pages}{204304} (\bibinfo{year}{2013}).

\bibitem[{\citenamefont{Perdew et~al.}(1996)\citenamefont{Perdew, Burke, and
  Ernzerhof}}]{PerBurErn-PRL-96}
\bibinfo{author}{\bibfnamefont{J.~P.} \bibnamefont{Perdew}},
  \bibinfo{author}{\bibfnamefont{K.}~\bibnamefont{Burke}}, \bibnamefont{and}
  \bibinfo{author}{\bibfnamefont{M.}~\bibnamefont{Ernzerhof}},
  \bibinfo{journal}{Phys. Rev. Lett.} \textbf{\bibinfo{volume}{77}},
  \bibinfo{pages}{3865} (\bibinfo{year}{1996}).

\bibitem[{\citenamefont{Perdew et~al.}(2008)\citenamefont{Perdew, Ruzsinszky,
  Csonka, Vydrov, Scuseria, Constantin, Zhou, and
  Burke}}]{PerRuzCsoVydScuConZhoBur-PRL-08}
\bibinfo{author}{\bibfnamefont{J.~P.} \bibnamefont{Perdew}},
  \bibinfo{author}{\bibfnamefont{A.}~\bibnamefont{Ruzsinszky}},
  \bibinfo{author}{\bibfnamefont{G.~I.} \bibnamefont{Csonka}},
  \bibinfo{author}{\bibfnamefont{O.~A.} \bibnamefont{Vydrov}},
  \bibinfo{author}{\bibfnamefont{G.~E.} \bibnamefont{Scuseria}},
  \bibinfo{author}{\bibfnamefont{L.~A.} \bibnamefont{Constantin}},
  \bibinfo{author}{\bibfnamefont{X.}~\bibnamefont{Zhou}}, \bibnamefont{and}
  \bibinfo{author}{\bibfnamefont{K.}~\bibnamefont{Burke}},
  \bibinfo{journal}{Phys. Rev. Lett.} \textbf{\bibinfo{volume}{100}},
  \bibinfo{pages}{136406} (\bibinfo{year}{2008}).

\bibitem[{\citenamefont{Sharkas et~al.}(2011)\citenamefont{Sharkas, Toulouse,
  and Savin}}]{ShaTouSav-JCP-11}
\bibinfo{author}{\bibfnamefont{K.}~\bibnamefont{Sharkas}},
  \bibinfo{author}{\bibfnamefont{J.}~\bibnamefont{Toulouse}}, \bibnamefont{and}
  \bibinfo{author}{\bibfnamefont{A.}~\bibnamefont{Savin}}, \bibinfo{journal}{J.
  Chem. Phys.} \textbf{\bibinfo{volume}{134}}, \bibinfo{pages}{064113}
  (\bibinfo{year}{2011}).

\bibitem[{\citenamefont{Zhao et~al.}(2004)\citenamefont{Zhao, Lynch, and
  Truhlar}}]{ZhaLynTru-JPCA-04}
\bibinfo{author}{\bibfnamefont{Y.}~\bibnamefont{Zhao}},
  \bibinfo{author}{\bibfnamefont{B.~J.} \bibnamefont{Lynch}}, \bibnamefont{and}
  \bibinfo{author}{\bibfnamefont{D.~G.} \bibnamefont{Truhlar}},
  \bibinfo{journal}{J. Phys. Chem. A} \textbf{\bibinfo{volume}{108}},
  \bibinfo{pages}{4786} (\bibinfo{year}{2004}).

\bibitem[{\citenamefont{Zhao et~al.}(2005)\citenamefont{Zhao, Lynch, and
  Truhlar}}]{ZhaLynTru-PCCP-05}
\bibinfo{author}{\bibfnamefont{Y.}~\bibnamefont{Zhao}},
  \bibinfo{author}{\bibfnamefont{B.~J.} \bibnamefont{Lynch}}, \bibnamefont{and}
  \bibinfo{author}{\bibfnamefont{D.~G.} \bibnamefont{Truhlar}},
  \bibinfo{journal}{Phys. Chem. Chem. Phys.} \textbf{\bibinfo{volume}{7}},
  \bibinfo{pages}{43} (\bibinfo{year}{2005}).

\bibitem[{\citenamefont{Peverati and Head-Gordon}(2013)}]{PevHea-JCP-13}
\bibinfo{author}{\bibfnamefont{R.}~\bibnamefont{Peverati}} \bibnamefont{and}
  \bibinfo{author}{\bibfnamefont{M.}~\bibnamefont{Head-Gordon}},
  \bibinfo{journal}{J. Chem. Phys.} \textbf{\bibinfo{volume}{139}},
  \bibinfo{pages}{024110} (\bibinfo{year}{2013}).

\bibitem[{\citenamefont{Schwabe and Grimme}(2006)}]{SchGri-PCCP-06}
\bibinfo{author}{\bibfnamefont{T.}~\bibnamefont{Schwabe}} \bibnamefont{and}
  \bibinfo{author}{\bibfnamefont{S.}~\bibnamefont{Grimme}},
  \bibinfo{journal}{Phys. Chem. Chem. Phys.} \textbf{\bibinfo{volume}{8}},
  \bibinfo{pages}{4398} (\bibinfo{year}{2006}).

\bibitem[{\citenamefont{Tarnopolsky et~al.}(2008)\citenamefont{Tarnopolsky,
  Karton, Sertchook, Vuzman, and Martin}}]{TarKarSerVuzMar-JPCA-08}
\bibinfo{author}{\bibfnamefont{A.}~\bibnamefont{Tarnopolsky}},
  \bibinfo{author}{\bibfnamefont{A.}~\bibnamefont{Karton}},
  \bibinfo{author}{\bibfnamefont{R.}~\bibnamefont{Sertchook}},
  \bibinfo{author}{\bibfnamefont{D.}~\bibnamefont{Vuzman}}, \bibnamefont{and}
  \bibinfo{author}{\bibfnamefont{J.~M.~L.} \bibnamefont{Martin}},
  \bibinfo{journal}{J. Phys. Chem. A} \textbf{\bibinfo{volume}{112}},
  \bibinfo{pages}{3} (\bibinfo{year}{2008}).

\bibitem[{\citenamefont{Karton et~al.}(2008)\citenamefont{Karton, Tarnopolsky,
  Lam\`ere, Schatz, and Martin}}]{KarTarLamSchMar-JPCA-08}
\bibinfo{author}{\bibfnamefont{A.}~\bibnamefont{Karton}},
  \bibinfo{author}{\bibfnamefont{A.}~\bibnamefont{Tarnopolsky}},
  \bibinfo{author}{\bibfnamefont{J.-F.} \bibnamefont{Lam\`ere}},
  \bibinfo{author}{\bibfnamefont{G.~C.} \bibnamefont{Schatz}},
  \bibnamefont{and} \bibinfo{author}{\bibfnamefont{J.~M.~L.}
  \bibnamefont{Martin}}, \bibinfo{journal}{J. Phys. Chem. A}
  \textbf{\bibinfo{volume}{112}}, \bibinfo{pages}{12868}
  (\bibinfo{year}{2008}).

\bibitem[{\citenamefont{Sancho-Garc\'ia and
  P\'erez-Jim\'enez}(2009)}]{SanPer-JCP-09}
\bibinfo{author}{\bibfnamefont{J.~C.} \bibnamefont{Sancho-Garc\'ia}}
  \bibnamefont{and} \bibinfo{author}{\bibfnamefont{A.~J.}
  \bibnamefont{P\'erez-Jim\'enez}}, \bibinfo{journal}{J. Chem. Phys.}
  \textbf{\bibinfo{volume}{131}}, \bibinfo{pages}{084108}
  (\bibinfo{year}{2009}).

\bibitem[{\citenamefont{Curtiss et~al.}(1997)\citenamefont{Curtiss,
  Raghavachari, Redfern, and Pople}}]{CurRagRedPop-JCP-97}
\bibinfo{author}{\bibfnamefont{L.~A.} \bibnamefont{Curtiss}},
  \bibinfo{author}{\bibfnamefont{K.}~\bibnamefont{Raghavachari}},
  \bibinfo{author}{\bibfnamefont{P.~C.} \bibnamefont{Redfern}},
  \bibnamefont{and} \bibinfo{author}{\bibfnamefont{J.~A.} \bibnamefont{Pople}},
  \bibinfo{journal}{J. Chem. Phys.} \textbf{\bibinfo{volume}{106}},
  \bibinfo{pages}{1063} (\bibinfo{year}{1997}).

\bibitem[{\citenamefont{Toulouse et~al.}(2004)\citenamefont{Toulouse, Colonna,
  and Savin}}]{TouColSav-PRA-04}
\bibinfo{author}{\bibfnamefont{J.}~\bibnamefont{Toulouse}},
  \bibinfo{author}{\bibfnamefont{F.}~\bibnamefont{Colonna}}, \bibnamefont{and}
  \bibinfo{author}{\bibfnamefont{A.}~\bibnamefont{Savin}},
  \bibinfo{journal}{Phys. Rev. A} \textbf{\bibinfo{volume}{70}},
  \bibinfo{pages}{062505} (\bibinfo{year}{2004}).

\bibitem[{\citenamefont{\'Angy\'an et~al.}(2005)\citenamefont{\'Angy\'an,
  Gerber, Savin, and Toulouse}}]{AngGerSavTou-PRA-05}
\bibinfo{author}{\bibfnamefont{J.~G.} \bibnamefont{\'Angy\'an}},
  \bibinfo{author}{\bibfnamefont{I.~C.} \bibnamefont{Gerber}},
  \bibinfo{author}{\bibfnamefont{A.}~\bibnamefont{Savin}}, \bibnamefont{and}
  \bibinfo{author}{\bibfnamefont{J.}~\bibnamefont{Toulouse}},
  \bibinfo{journal}{Phys. Rev. A} \textbf{\bibinfo{volume}{72}},
  \bibinfo{pages}{012510} (\bibinfo{year}{2005}).

\bibitem[{\citenamefont{Fromager and Jensen}(2008)}]{FroJen-PRA-08}
\bibinfo{author}{\bibfnamefont{E.}~\bibnamefont{Fromager}} \bibnamefont{and}
  \bibinfo{author}{\bibfnamefont{H.~J.~A.} \bibnamefont{Jensen}},
  \bibinfo{journal}{Phys. Rev. A} \textbf{\bibinfo{volume}{78}},
  \bibinfo{pages}{022504} (\bibinfo{year}{2008}).

\bibitem[{\citenamefont{\'Angy\'an}(2008)}]{Ang-PRA-08}
\bibinfo{author}{\bibfnamefont{J.~G.} \bibnamefont{\'Angy\'an}},
  \bibinfo{journal}{Phys. Rev. A} \textbf{\bibinfo{volume}{78}},
  \bibinfo{pages}{022510} (\bibinfo{year}{2008}).

\bibitem[{\citenamefont{Pisani et~al.}(2008)\citenamefont{Pisani, Maschio,
  Casassa, Halo, Sch\"utz, and Usvyat}}]{PisMasCasHalSchUsv-JCC-08}
\bibinfo{author}{\bibfnamefont{C.}~\bibnamefont{Pisani}},
  \bibinfo{author}{\bibfnamefont{L.}~\bibnamefont{Maschio}},
  \bibinfo{author}{\bibfnamefont{S.}~\bibnamefont{Casassa}},
  \bibinfo{author}{\bibfnamefont{M.}~\bibnamefont{Halo}},
  \bibinfo{author}{\bibfnamefont{M.}~\bibnamefont{Sch\"utz}}, \bibnamefont{and}
  \bibinfo{author}{\bibfnamefont{D.}~\bibnamefont{Usvyat}},
  \bibinfo{journal}{J. Comput. Chem.} \textbf{\bibinfo{volume}{29}},
  \bibinfo{pages}{2113} (\bibinfo{year}{2008}).

\bibitem[{\citenamefont{Pulay and Saeb{\o}}(1986)}]{PulSae-TCA-86}
\bibinfo{author}{\bibfnamefont{P.}~\bibnamefont{Pulay}} \bibnamefont{and}
  \bibinfo{author}{\bibfnamefont{S.}~\bibnamefont{Saeb{\o}}},
  \bibinfo{journal}{Theor. Chim. Acta} \textbf{\bibinfo{volume}{69}},
  \bibinfo{pages}{357} (\bibinfo{year}{1986}).

\bibitem[{\citenamefont{{R. P. Steele and R. A. DiStasio and Y. Shao and J.
  Kong and M. Head-Gordon}}(2006)}]{SteDiSShaKonHea-JCP-06}
\bibinfo{author}{\bibnamefont{{R. P. Steele and R. A. DiStasio and Y. Shao and
  J. Kong and M. Head-Gordon}}}, \bibinfo{journal}{J. Chem. Phys.}
  \textbf{\bibinfo{volume}{125}}, \bibinfo{pages}{074108}
  (\bibinfo{year}{2006}).

\bibitem[{\citenamefont{{R. A. DiStasio and R. P. Steele and M.
  Head-Gordon}}(2007)}]{DiSSteHea-MP-07}
\bibinfo{author}{\bibnamefont{{R. A. DiStasio and R. P. Steele and M.
  Head-Gordon}}}, \bibinfo{journal}{Mol. Phys.} \textbf{\bibinfo{volume}{105}},
  \bibinfo{pages}{2731} (\bibinfo{year}{2007}).

\bibitem[{\citenamefont{Usvyat et~al.}(2010)\citenamefont{Usvyat, Maschio,
  Pisani, and Sch\"utz}}]{UsvMasPisSch-ZPC-10}
\bibinfo{author}{\bibfnamefont{D.}~\bibnamefont{Usvyat}},
  \bibinfo{author}{\bibfnamefont{L.}~\bibnamefont{Maschio}},
  \bibinfo{author}{\bibfnamefont{C.}~\bibnamefont{Pisani}}, \bibnamefont{and}
  \bibinfo{author}{\bibfnamefont{M.}~\bibnamefont{Sch\"utz}},
  \bibinfo{journal}{Z. Phys. Chem.} \textbf{\bibinfo{volume}{224}},
  \bibinfo{pages}{441} (\bibinfo{year}{2010}).

\bibitem[{\citenamefont{Dovesi et~al.}()\citenamefont{Dovesi, Saunders, Roetti,
  Orlando, Zicovich-Wilson, Pascale, Civalleri, Doll, Harrison, Bush
  et~al.}}]{Cry-PROG-09}
\bibinfo{author}{\bibfnamefont{R.}~\bibnamefont{Dovesi}},
  \bibinfo{author}{\bibfnamefont{V.}~\bibnamefont{Saunders}},
  \bibinfo{author}{\bibfnamefont{C.}~\bibnamefont{Roetti}},
  \bibinfo{author}{\bibfnamefont{R.}~\bibnamefont{Orlando}},
  \bibinfo{author}{\bibfnamefont{C.}~\bibnamefont{Zicovich-Wilson}},
  \bibinfo{author}{\bibfnamefont{F.}~\bibnamefont{Pascale}},
  \bibinfo{author}{\bibfnamefont{B.}~\bibnamefont{Civalleri}},
  \bibinfo{author}{\bibfnamefont{K.}~\bibnamefont{Doll}},
  \bibinfo{author}{\bibfnamefont{N.}~\bibnamefont{Harrison}},
  \bibinfo{author}{\bibfnamefont{I.}~\bibnamefont{Bush}}, \bibnamefont{et~al.},
  \emph{\bibinfo{title}{Crystal, release crystal09 (2009), see
  http://www.crystal.unito.it}}.

\bibitem[{\citenamefont{Allen}(2002)}]{All-AC-02}
\bibinfo{author}{\bibfnamefont{F.~H.} \bibnamefont{Allen}},
  \bibinfo{journal}{Acta Cryst.} \textbf{\bibinfo{volume}{B58}},
  \bibinfo{pages}{380} (\bibinfo{year}{2002}).

\bibitem[{\citenamefont{Simon and Peters}(1980)}]{SimPet-AC-80}
\bibinfo{author}{\bibfnamefont{A.}~\bibnamefont{Simon}} \bibnamefont{and}
  \bibinfo{author}{\bibfnamefont{K.}~\bibnamefont{Peters}},
  \bibinfo{journal}{Acta Cryst.} \textbf{\bibinfo{volume}{B36}},
  \bibinfo{pages}{2750} (\bibinfo{year}{1980}).

\bibitem[{\citenamefont{Kiefte et~al.}(1987)\citenamefont{Kiefte, Penney,
  Breckon, and Clouter}}]{KiePenBreClo-JCP-87}
\bibinfo{author}{\bibfnamefont{H.}~\bibnamefont{Kiefte}},
  \bibinfo{author}{\bibfnamefont{R.}~\bibnamefont{Penney}},
  \bibinfo{author}{\bibfnamefont{S.~W.} \bibnamefont{Breckon}},
  \bibnamefont{and} \bibinfo{author}{\bibfnamefont{M.~J.}
  \bibnamefont{Clouter}}, \bibinfo{journal}{J. Chem. Phys.}
  \textbf{\bibinfo{volume}{86}}, \bibinfo{pages}{662} (\bibinfo{year}{1987}).

\bibitem[{\citenamefont{Voitovich et~al.}(1971)\citenamefont{Voitovich,
  Tolkachev, and Manzhelii}}]{VoiTolMan-JLTP-71}
\bibinfo{author}{\bibfnamefont{E.~I.} \bibnamefont{Voitovich}},
  \bibinfo{author}{\bibfnamefont{A.~M.} \bibnamefont{Tolkachev}},
  \bibnamefont{and} \bibinfo{author}{\bibfnamefont{V.~G.}
  \bibnamefont{Manzhelii}}, \bibinfo{journal}{J. Low Temp. Phys.}
  \textbf{\bibinfo{volume}{5}}, \bibinfo{pages}{435} (\bibinfo{year}{1971}).

\bibitem[{\citenamefont{Willock et~al.}(1995)\citenamefont{Willock, Price,
  Leslie, and Catlow}}]{WilPriLesCat-JCC-95}
\bibinfo{author}{\bibfnamefont{D.~J.} \bibnamefont{Willock}},
  \bibinfo{author}{\bibfnamefont{S.~L.} \bibnamefont{Price}},
  \bibinfo{author}{\bibfnamefont{M.}~\bibnamefont{Leslie}}, \bibnamefont{and}
  \bibinfo{author}{\bibfnamefont{C.~R.~A.} \bibnamefont{Catlow}},
  \bibinfo{journal}{J. Comput. Chem} \textbf{\bibinfo{volume}{16}},
  \bibinfo{pages}{628} (\bibinfo{year}{1995}).

\bibitem[{\citenamefont{Gavezzotti}(2008)}]{Gav-MP-08}
\bibinfo{author}{\bibfnamefont{A.}~\bibnamefont{Gavezzotti}},
  \bibinfo{journal}{Mol. Phys.} \textbf{\bibinfo{volume}{106}},
  \bibinfo{pages}{1473} (\bibinfo{year}{2008}).

\bibitem[{\citenamefont{Hariharan and Pople}(28)}]{HarPop-TCA-73}
\bibinfo{author}{\bibfnamefont{P.}~\bibnamefont{Hariharan}} \bibnamefont{and}
  \bibinfo{author}{\bibfnamefont{J.}~\bibnamefont{Pople}},
  \bibinfo{journal}{Theoret. Chim. Acta} \textbf{\bibinfo{volume}{1973}},
  \bibinfo{pages}{213} (\bibinfo{year}{28}).

\bibitem[{\citenamefont{Perger et~al.}(2004)\citenamefont{Perger, Pandey,
  Blanco, and Zhao}}]{PerPanBlaZha-CPL-04}
\bibinfo{author}{\bibfnamefont{W.~F.} \bibnamefont{Perger}},
  \bibinfo{author}{\bibfnamefont{R.}~\bibnamefont{Pandey}},
  \bibinfo{author}{\bibfnamefont{M.~A.} \bibnamefont{Blanco}},
  \bibnamefont{and} \bibinfo{author}{\bibfnamefont{J.}~\bibnamefont{Zhao}},
  \bibinfo{journal}{Chem. Phys. Lett.} \textbf{\bibinfo{volume}{{388}}},
  \bibinfo{pages}{175} (\bibinfo{year}{2004}).

\bibitem[{\citenamefont{Dunning}(1989)}]{Dun-JCP-89}
\bibinfo{author}{\bibfnamefont{T.~H.} \bibnamefont{Dunning}},
  \bibinfo{journal}{J. Chem. Phys.} \textbf{\bibinfo{volume}{90}},
  \bibinfo{pages}{1007} (\bibinfo{year}{1989}).

\bibitem[{\citenamefont{Zicovich-Wilson
  et~al.}(2001)\citenamefont{Zicovich-Wilson, Dovesi, and
  Saunders}}]{ZicDovSau-JCP-05}
\bibinfo{author}{\bibfnamefont{C.}~\bibnamefont{Zicovich-Wilson}},
  \bibinfo{author}{\bibfnamefont{R.}~\bibnamefont{Dovesi}}, \bibnamefont{and}
  \bibinfo{author}{\bibfnamefont{V.~R.} \bibnamefont{Saunders}},
  \bibinfo{journal}{J. Chem. Phys.} \textbf{\bibinfo{volume}{115}},
  \bibinfo{pages}{9708} (\bibinfo{year}{2001}).

\bibitem[{\citenamefont{Casassa et~al.}(2006)\citenamefont{Casassa,
  Zicovich-Wilson, and Pisani}}]{CasZicPia-TCA-06}
\bibinfo{author}{\bibfnamefont{S.}~\bibnamefont{Casassa}},
  \bibinfo{author}{\bibfnamefont{C.}~\bibnamefont{Zicovich-Wilson}},
  \bibnamefont{and} \bibinfo{author}{\bibfnamefont{C.}~\bibnamefont{Pisani}},
  \bibinfo{journal}{Theor. Chem. Acc.} \textbf{\bibinfo{volume}{116}},
  \bibinfo{pages}{726} (\bibinfo{year}{2006}).

\bibitem[{\citenamefont{Maschio and Usvyat}(2008)}]{MasUsv-PRB-08}
\bibinfo{author}{\bibfnamefont{L.}~\bibnamefont{Maschio}} \bibnamefont{and}
  \bibinfo{author}{\bibfnamefont{D.}~\bibnamefont{Usvyat}},
  \bibinfo{journal}{Phys. Rev. B} \textbf{\bibinfo{volume}{{78}}},
  \bibinfo{pages}{073102} (\bibinfo{year}{2008}).

\bibitem[{\citenamefont{Boys and Bernardi}(1970)}]{BoyBer-MP-70}
\bibinfo{author}{\bibfnamefont{S.~F.} \bibnamefont{Boys}} \bibnamefont{and}
  \bibinfo{author}{\bibfnamefont{F.}~\bibnamefont{Bernardi}},
  \bibinfo{journal}{Mol. Phys.} \textbf{\bibinfo{volume}{{19}}},
  \bibinfo{pages}{553} (\bibinfo{year}{1970}).

\bibitem[{Nis()}]{Nis-BOOK-13}
\emph{\bibinfo{title}{{NIST WebBook (The National Institute of Standards and
  Technology)}}}, \urlprefix\url{http://webbook.nist.gov}.

\bibitem[{\citenamefont{{A. Reilly and A. Tkatchenko}}(2012)}]{ReiTka-JCP-13}
\bibinfo{author}{\bibnamefont{{A. Reilly and A. Tkatchenko}}},
  \bibinfo{journal}{J. Chem. Phys.} \textbf{\bibinfo{volume}{139}},
  \bibinfo{pages}{024705} (\bibinfo{year}{2012}).

\bibitem[{\citenamefont{Chickos}(2003)}]{Chi-NS-03}
\bibinfo{author}{\bibfnamefont{J.~S.} \bibnamefont{Chickos}},
  \bibinfo{journal}{Netsu Sokutei} \textbf{\bibinfo{volume}{30}},
  \bibinfo{pages}{125} (\bibinfo{year}{2003}).

\bibitem[{\citenamefont{{H.-J. Werner, P. J. Knowles, G. Knizia, F. R. Manby,
  M. {Sch\"{u}tz}, and others}}()}]{Molproshort-PROG-12}
\bibinfo{author}{\bibnamefont{{H.-J. Werner, P. J. Knowles, G. Knizia, F. R.
  Manby, M. {Sch\"{u}tz}, and others}}}, \emph{\bibinfo{title}{Molpro, version
  2012.1, a package of ab initio programs}}, \bibinfo{note}{cardiff, UK, 2012,
  see \url{http://www.molpro.net}}.

\bibitem[{\citenamefont{Karton and Martin}(2011)}]{KarMar-JCP-11}
\bibinfo{author}{\bibfnamefont{A.}~\bibnamefont{Karton}} \bibnamefont{and}
  \bibinfo{author}{\bibfnamefont{J.~M.~L.} \bibnamefont{Martin}},
  \bibinfo{journal}{J. Chem. Phys.} \textbf{\bibinfo{volume}{135}},
  \bibinfo{pages}{144119} (\bibinfo{year}{2011}).

\bibitem[{\citenamefont{Chuang and Chen}(2011)}]{ChuChe-JCC-11}
\bibinfo{author}{\bibfnamefont{Y.-Y.} \bibnamefont{Chuang}} \bibnamefont{and}
  \bibinfo{author}{\bibfnamefont{S.-M.} \bibnamefont{Chen}},
  \bibinfo{journal}{J. Comput. Chem.} \textbf{\bibinfo{volume}{32}},
  \bibinfo{pages}{1671} (\bibinfo{year}{2011}).

\bibitem[{\citenamefont{Grimme et~al.}(2010)\citenamefont{Grimme, Antony,
  Ehrlich, and Krieg}}]{GriAntEhrKri-JCP-10}
\bibinfo{author}{\bibfnamefont{S.}~\bibnamefont{Grimme}},
  \bibinfo{author}{\bibfnamefont{J.}~\bibnamefont{Antony}},
  \bibinfo{author}{\bibfnamefont{S.}~\bibnamefont{Ehrlich}}, \bibnamefont{and}
  \bibinfo{author}{\bibfnamefont{H.}~\bibnamefont{Krieg}}, \bibinfo{journal}{J.
  Chem. Phys.} \textbf{\bibinfo{volume}{132}}, \bibinfo{pages}{154104}
  (\bibinfo{year}{2010}).

\bibitem[{\citenamefont{Müller and Usvyat}(2013)}]{MUsv}
\bibinfo{author}{\bibfnamefont{C.}~\bibnamefont{Müller}} \bibnamefont{and}
  \bibinfo{author}{\bibfnamefont{D.}~\bibnamefont{Usvyat}},
  \bibinfo{journal}{J. Chem. Theory Comput.} \textbf{\bibinfo{volume}{9}},
  \bibinfo{pages}{5590} (\bibinfo{year}{2013}).

\bibitem[{\citenamefont{Chickos and Acree}(2002)}]{ChiAcr-JPCRD-02}
\bibinfo{author}{\bibfnamefont{J.~S.} \bibnamefont{Chickos}} \bibnamefont{and}
  \bibinfo{author}{\bibfnamefont{W.~E.} \bibnamefont{Acree}},
  \bibinfo{journal}{J. Phys. Chem. Ref. Data.} \textbf{\bibinfo{volume}{31}},
  \bibinfo{pages}{537} (\bibinfo{year}{2002}).

\bibitem[{\citenamefont{Paulus}(2006)}]{Paulus:2006p14134}
\bibinfo{author}{\bibfnamefont{B.}~\bibnamefont{Paulus}},
  \bibinfo{journal}{Phys. Rep.} \textbf{\bibinfo{volume}{428}},
  \bibinfo{pages}{1} (\bibinfo{year}{2006}).

\end{thebibliography}
\end{document}